\documentclass{svjour3}
\usepackage[utf8]{inputenc}
\usepackage[russian,english]{babel}
\usepackage[crossref=false]{pstool}
\usepackage{bm}
\usepackage{amssymb}
\usepackage{amsxtra}
\usepackage[mathscr]{eucal}

\usepackage[svgnames]{xcolor}
\usepackage{graphicx}

\EndPreamble
\usepackage{mathtools}
\usepackage{bigints}

\usepackage[misc]{ifsym}

\usepackage[colorlinks,allcolors=blue,unicode]{hyperref}
\usepackage{cite}

\hypersetup{
    pdftitle = {Ballistic thermal transport in graphene lattice}
}

\makeatletter

\@addtoreset{equation}{section}
\makeatother
\journalname{CMAT}

\def\I{\mathrm{i}}
\def\kappaT{T}

\def\breveC{{\mathscr C_0}}

\def\be#1{\begin{equation}\label{#1}}
\def\ee{\end{equation}}
\newcommand {\ba}[2]{\be{#1}\begin{array}{#2}}
\newcommand {\ea}{\end{array} \ee}

\renewcommand{\=}{\stackrel{\mbox{\scriptsize def}}{=}}

\def\({\left(}
\def\){\right)}

\def\dt{\partial_t}

\let\de = \delta

\def\d{{\cal D}}

\def\qy{{y}}

\def\chio{\bar\chi_0}

\def\pd#1#2{\frac{\partial #1}{\partial #2}}

\def\bb#1{\mathbf #1}
\def\d{\mathrm d}

\def\bC{{\mathbf g}}

\def\groopv{{\mathbf g}}
\def\Groopv{{\mathbf G}}

\begin{document}

\selectlanguage{english}
\author{Serge N.~Gavrilov \and Anton M.~Krivtsov}
\title{Steady-state ballistic thermal transport associated with transversal
motions in a damped graphene lattice 
subjected to a point heat source%
}
\titlerunning{Ballistic thermal transport in graphene lattice}
%
\institute{
S.N.~Gavrilov (\Letter) \at
Institute for Problems in Mechanical Engineering RAS, V.O., Bolshoy pr.~61,
St.~Petersburg, 199178, Russia \\
\email{serge@pdmi.ras.ru}           
\and
A.M.~Krivtsov \at
Peter the Great St.~Petersburg Polytechnic University (SPbPU),
Polytechnicheskaya str.~29, St.Petersburg, 195251, Russia\\
\email{akrivtsov@bk.ru}
\and
A.M.~Krivtsov \at
Institute for Problems in Mechanical Engineering RAS, V.O., Bolshoy pr.~61,
St.~Petersburg, 199178, Russia}
%

\maketitle


%
%
\def\tr{\operatorname{tr}}
\def\breveC{\mathscr A}


\begin{abstract} 
In the paper we deal with ballistic heat transport in a graphene lattice
subjected to a point heat source.   It is assumed that a graphene sheet is
suspended under tension in a viscous gas.  We use the model of a harmonic
polyatomic (more exactly diatomic) lattice performing out-of-plane motions.
The dynamics of the lattice is described by an infinite system
of stochastic ordinary differential equations with white noise in the
right-hand side, which models the point heat source.
On the base of the previous analytical unsteady analysis an analytical formula
in continuum approximation is suggested, which allows one to
describe a steady-state kinetic temperature distribution in the graphene
lattice in continuum approximation. The obtained solution is in a good agreement with numerical results
obtained for the discrete system everywhere excepting a neighbourhood of
six singular rays with the origin at the heat source location. The continuum
solution becomes singular at these rays, unlike the discrete one, which appears to
be localized in a certain sense along the rays. The factors, which cause such a
directional localization and the mismatch between the continuum and discrete
solutions are discussed. We expect that the suggested formula is applicable for
various damped polyatomic lattices where all particles have
equal masses in
the case of universal for all particles external viscosity.  
\end{abstract}


\section{Introduction}
\label{gra-sec-intro}

Analytical studies of the heat transfer in low-dimensional lattices have
demonstrated that the classical Fourier law in such systems is frequently
violated and can be substituted by non-classical types of thermal behaviour 
\cite{rieder1967properties,lepri2003thermal,dhar2015heat,savin2016normal}.
Recent experimental investigations of the thermal transport in graphene indicate that under certain
conditions the thermal conductivity can be size dependent, and, in particular,
ballistic \cite{Nika_2017,Bae2013,Xu2014,Serov_2013}.
The present paper is a theoretic study of ballistic heat transport in a graphene lattice
subjected to a point heat source. It is assumed that a graphene sheet is
suspended under tension in a viscous gas.  We use the model of a harmonic
polyatomic (more exactly diatomic) lattice performing out-of-plane motions.
A harmonic lattice is a very simple mechanical model that allows one to obtain
solutions of complicated problems in analytical form. Since the pioneering
studies (see \cite{rieder1967properties}) it is known that this model describes
ballistic thermal conductivity.
The dynamics of the lattice is described by an infinite system
of stochastic ordinary differential equations with white noise in the
right-hand side, which models the heat source.
The aim of the paper is to obtain the approximate continuum
solution, which describes the steady-state kinetic temperature distribution in
the lattice caused by the point heat source of constant intensity.  Note that
we need some damping (a gas) for the existence of a limiting steady-state solution.
Despite the fact that our investigation is theoretical, we indicate that
we have chosen such a problem keeping in mind that the solution, apparently, can be
compared with results of experiments with pure monocrystalline
graphene. 

To get the analytical formula describing the steady-state kinetic temperature
distribution,  we
implement informal generalization of results of previous
papers 
\cite{gavrilov2018heat,gavrilov2019steady,Kuzkin-Krivtsov-accepted,Kuzkin2019}
concerning ballistic heat transport in primitive scalar
\cite{gavrilov2018heat,gavrilov2019steady,Kuzkin-Krivtsov-accepted}
and polyatomic 
\cite{Kuzkin2019,kuzkin2019thermal} 
lattices.
The approach used in studies 
\cite{gavrilov2018heat,gavrilov2019steady,Kuzkin-Krivtsov-accepted}
is suggested in
\cite{krivtsov2015heat}
and based on introducing and dealing with 
infinite set of covariance variables in spirit of \cite{rieder1967properties}. These  
are the mutual covariances of all the particle velocities and all the
displacements for all pairs of particles. Applying the It\^o lemma allows one
to derive an infinite deterministic system of ordinary differential equations
which follows from the equations of stochastic dynamics. Then 
{the procedure of
continualization is applied} to rewrite the finite difference operators
involved in the equation for covariances as compositions of finite difference
operators and operators of differentiation. 
The next step is the separation of slow motions, which are related to heat
propagation (the fast motions are energy oscillation associated with the
transformation of the kinetic energy to the potential one and in backward
direction, i.e.\ the thermal equilibration
\cite{hemmer1959dynamic,
klein1953mecanique,krivtsov2014energy,kuzkin2019thermal,Gavrilov2019,Sokolov2021,Berinskii2020}). This
approach is very efficient for scalar lattices,
however, our attempts to apply it to polyatomic lattices were not
successful yet.

{For polyatomic lattices 
in \cite{Kuzkin2019} a different technique based on the direct solution of the dynamics equations is used 
to evaluate a continuum approximation for the kinetic temperature field.
At the
first step, an exact expression for the matrix of covariances of particle
velocities inside a primitive cell (the temperature matrix)
in the form of a multiple integral has to be found. Note that the conventional kinetic
temperature for a cell is the trace of the temperature matrix.
At the second step, several integrals are evaluated using an approximate
procedure, which is related, apparently, to the asymptotic method of stationary phase
\cite{Fedoruk-Saddle,temme2014}.
Finally, the kinetic temperature is found as the sum of a slow continuum component and
a fast one.} Note that in the case of primitive scalar lattices both
approaches lead to the same result.

Both of two approaches discussed above were verified by numerical calculations
based on discrete equations of stochastic dynamics and  
performed for various lattices.
An excellent agreement was demonstrated.
In particular, in \cite{Kuzkin2019} non-stationary ballistic
heat transport in graphene lattice caused by an impulse point source 
in the undamped (conservative) case is considered.
However, in comparison with the first one the approach used in 
\cite{Kuzkin2019}
have two limitations. One of them 
is related to the fact that in \cite{Kuzkin2019} the heat sources are
considered only in the form of random initial conditions for particle
velocities. This allows one to deal with commonly used (non-stochastic) ODE to 
formulate the
problem.
The heat sources that act after the initial instant of time are
beyond the scope of the paper \cite{Kuzkin2019}.\footnote{To describe such sources
one needs to formulate a problem for a system of {\it stochastic} ordinary differential
equations.} The second limitation is
related to the fact that an external damping is not taken into account. In the
presence of the damping the method of stationary phase becomes inapplicable,
and we need to use the saddle-point method 
\cite{Fedoruk-Saddle,temme2014}.
Apparently, both limitations can be addressed, and the corresponding
procedure, which generalizes the second approach, can be suggested. However,
taking into account the damping essentially complicates (already quite
complicated) mathematical procedure, and for the time being we have not
finalized
such a work in general case. In the present paper, we instead guess the final
formula for approximate continuum solution in the particular case of damped
graphene lattice. We argue the possible
applicability of suggested formula 
basing on the structure of previously obtained in
\cite{gavrilov2018heat,gavrilov2019steady,Kuzkin-Krivtsov-accepted,Kuzkin2019}
particular results. Then
we verify the formula by numerical calculations based on the discrete model
and demonstrate a good agreement between the discrete solution and the
continuum one (Sect.~\ref{gra-sec-numerics}). The accurate derivation of the
suggested formula is beyond the scope of this paper. 

The paper is organized as follows. 
In Sect.~\ref{gra-sec-notation} we discuss the notation.
In Sect.~\ref{gra-sec-formulation} we
present the mathematical formulation for the discrete problem. In
Sect.~\ref{gra-sec-formulation-gra} we formulate equations of stochastic
dynamics for the graphene lattice.
Additionally, we need to consider an auxiliary problem related to
the primitive rhombic scalar lattice of graphene cell centres.
The corresponding mathematical formulation
is given in Sect.~\ref{gra-sec-rhombic-statement}.
In Sect.~\ref{sec-gra-approximate} we construct the approximate continuum
solution generalizing the results of papers 
\cite{gavrilov2018heat,gavrilov2019steady,Kuzkin-Krivtsov-accepted,Kuzkin2019}.
First, in Sect.~\ref{gra-sec-rhombic-sol}
we obtain the formula describing the steady-state kinetic
temperature distribution for the rhombic scalar lattice discussed 
in Sect.~\ref{gra-sec-rhombic-statement}. Then in Sect.~\ref{gra-sect-gra}
we suggest the corresponding formula for polyatomic graphene lattice in the
form of a double integral, and simplify its structure in Sect.~\ref{gra-sec-simplify}
transforming the double
integral into a single one. The continuum solution predicts the existence of
six rays with the origin at the point heat source location, where the continuum
solution is singular. In Sect.~\ref{gra-sec-singular} we discuss these rays
and corresponding physical phenomenon of ballistic phonon focusing in the
context of graphene.
In Sect.~\ref{gra-sec-numerics} we present the
results of the numerical solution of the initial value problem for the system
of stochastic differential equations and compare them with the obtained
continuum solution.
In the conclusion (Sect.~\ref{gra-sec-conclusion}) we discuss the basic
results of the paper. In Appendix~\ref{gra-sec-app-dispersion}
we provide formulas for 
the dispersion surfaces and the group velocities for graphene lattice,
obtained, e.g., in \cite{kuzkin2019thermal,Kuzkin2019}.

\section{Nomenclature}
\label{gra-sec-notation}
In the paper, we use the following general notation:
\begin{description}	
\item[$\mathbb Z$] is
the set of all integers;
\item[$\mathbb R$] is
the set of all real numbers; 
\item[$t$] is the time;
\item[$\dt$] is the differential operator with respect to time;
\item[$H(\cdot)$] is the Heaviside function;
\item[$\langle\cdot\rangle$] is the mathematical expectation for a random quantity;
\item[$\de_p^q\equiv\de_{pq}$ ] are {the Kronecker deltas}
({$p,q\in\mathbb Z$});
\item[$\bm I$] is the identity matrix;
\item[$k_B$] is the Boltzmann constant;
\item[$d$] is the lattice dimension ($d=2$ in the framework of the problems under
consideration);
\item[$Q$] is the number of cells with a primitive cell interacts.
\end{description}	

We use bold italic symbols for matrices 
and {bold upright}
symbols for invariant vectors in two-dimensional
space. The Einstein summation rule \cite{Kuptsov2001} is assumed. 
{To avoid any ambiguity, the accent $\,\breve{.}\,$ explicitly marks
discrete co-ordinates, invariant vectors with discrete co-ordinates, and
functions of invariant vectors with discrete co-ordinates in the cases where
the corresponding continuum quantities are used in the paper.}

\section{Mathematical formulation for the discrete problem}
\label{gra-sec-formulation}
\subsection{Graphene lattice}
\label{gra-sec-formulation-gra}

\begin{figure}[t]	
\centering\includegraphics[width=0.8\textwidth]{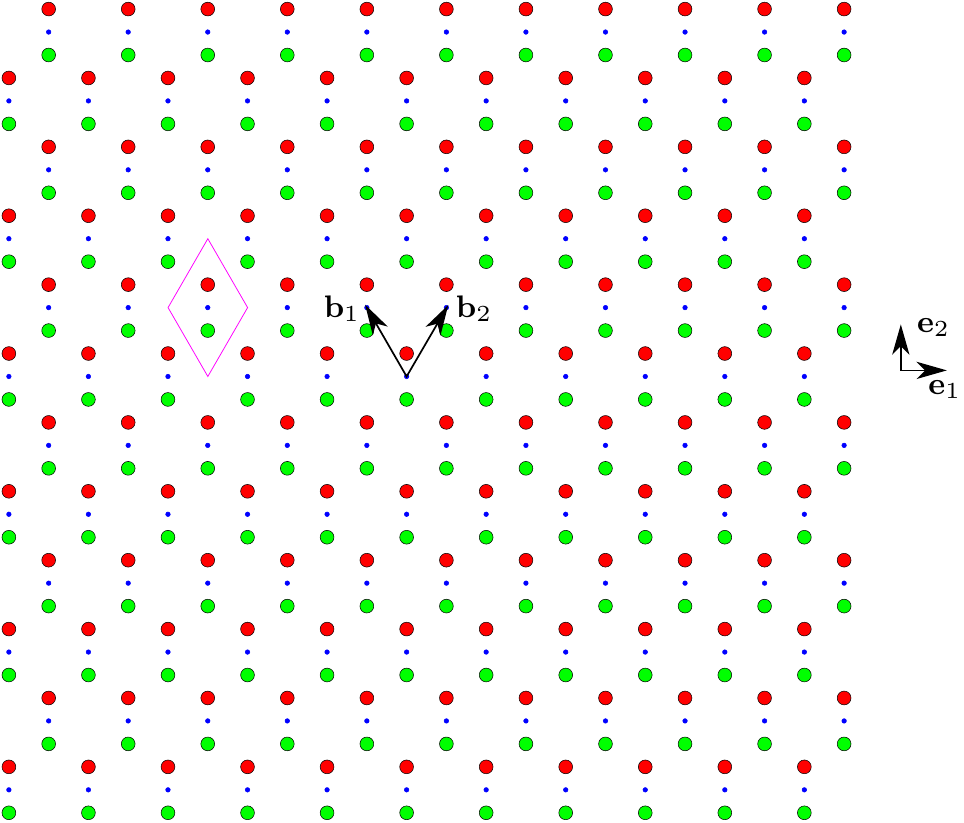}
\caption{Schematic of the graphene lattice: 
a primitive cell is shown by magenta colour, 
two material Bravais sub-lattices are shown by red and
green circles, the corresponding 
rhombic Bravais lattice of cell centres 
is shown by blue points
}
\label{graphene.pdf}
\end{figure}

We consider the transverse oscillation of an infinite graphene lattice. 
The schematic of the system is shown in Fig.~\ref{graphene.pdf}. 
There are 
\begin{equation}
N=2 
\label{gra-Nis2}
\end{equation}
particles in a primitive cell
(see Fig.~\ref{graphene.pdf}),
i.e.\ the particles are
arranged into two material sub-lattices, which have the same structure as 
the Bravais lattice \cite{kosevich2005the} of cells centres.
\begin{remark}  
{For better understanding of physical meaning and to keep general structure of
formulas, it is useful not to substitute variables $N$, $Q$, and $d$ by their
values, until the final result is obtained.  Moreover, in what follows 
(see Sect.~\ref{gra-sec-rhombic-statement}, \ref{gra-sec-rhombic-sol}), 
we also consider an auxiliary problem for the primitive rhombic lattice of graphene cell
centres, which corresponds to formal choice $N=1$}. 
\end{remark}
The masses $m$ of all the particles assumed to be equal.  
Every particle interacts with three neighbouring particles from the alternative
sub-lattice. 
These particles are located at the same distance%
\footnote{$a\simeq0.142$ nm for real graphene lattice \cite{Nika_2017}.} $a$ (in
equilibrium) that corresponds to the lattice constant $a_0=a\sqrt3$ for the
Bravais lattice.
One of the particles belongs to the same cell, and other two belong to
nearest-neighbour cells, thus a cell interacts with $Q=4$ cells.
Following \cite{kuzkin2019thermal,Kuzkin2019}, we use matrix notation to
formulate the basic equations of motion in the form of dynamic equations for a
polyatomic lattice \cite{kosevich2005the}. The column of displacements for $N=2$ particles in a primitive cell of the
graphene lattice is
\begin{equation}
\bm u (\breve{\mathbf x},t)=
\begin{pmatrix}	
u_1(\breve{\mathbf x},t) & \dots & u_N(\breve{\mathbf x},t)
\end{pmatrix}
^\top.
\label{gra-column-u}
\end{equation}
Here 
\begin{equation}
\breve{\bb x}=\breve x^\beta\bb b_\beta, \qquad \breve x^\beta\in\mathbb Z
\end{equation}
is the position vector for a cell centre,
$\bb b_\beta$ are the primitive vectors (see Fig.~\ref{graphene.pdf}):
\begin{equation}
\bb b_1=\frac{\sqrt3a}2(\bb e_1 +\sqrt3 \bb e_2 ),
\qquad
\bb b_2=\frac{\sqrt3a}2(-\bb e_1 +\sqrt3 \bb e_2 );
\label{gra-bbe}
\end{equation}
$\mathbf e_1$ and $\mathbf e_2$ are orthogonal unit vectors that correspond to
so-called zigzag and armchair directions, respectively (see Fig.~\ref{graphene.pdf}).
Note that according the Einstein summation rule \cite{Kuptsov2001}, which is used in the paper,
we use the following notation:
\begin{equation}
\breve x^\beta\bb b_\beta
\=
\sum_{\beta=1}^2\breve x^\beta\bb b_\beta.
\end{equation}

The stochastic equations of motion can be formulated as follows (see, e.g.,
\cite{Kuzkin2019}):
\begin{gather}
m\,\dt{\bm v}(\breve{\bb x})=-\eta_0 {\bm v}(\breve{\bb x})
-\bm C_0 \bm u(\breve{\bb x})
-\sum_{\beta=1}^{Q/2}
\left(
\bm C_1 \bm u(\breve{\bb x}+\bb b_\beta)
+\bm C_1^\top \bm u(\breve{\bb x}-\bb b_\beta)
\right)
+b_0(\breve{\bb x},t)
\,\dot{ \bm W, }
\label{gra-maineq}
\\
\dt{\bm u}(\breve{\bb x})=
\bm v(\breve{\bb x}).
\label{gra-v-dot-u}
\end{gather}
Here 
$\bm v$ is the column of the particle velocities; 
$\bm C_0$ and 
$\bm C_1$
are $N\times N$ stiffness matrices:
\begin{gather}
\bm C_0=
C\begin{pmatrix}
3 & -1 \\
-1 & 3
\end{pmatrix}
,
\qquad
\bm C_1=
C\begin{pmatrix}
0 & -1 \\
0 & 0
\end{pmatrix};
\label{gra-C-matrices}
\end{gather}
$C$ is the bond stiffness (tension between neighbour particles);
$b_0(\breve{\bb x},t)$ is the intensity of the random external excitation; 
\begin{equation}
\dot {\bm W}=
\begin{pmatrix}	
\dot {W_1}& \dots & \dot {W_N}
\end{pmatrix}^\top;
\end{equation}
{$W_i(\breve{\bb x})$ and $W_j(\breve{\bb y})$ are the uncorrelated Wiener
processes for $i\neq j$}
($\dot W_i(\breve{\bb x})$ and $\dot W_j(\breve{\bb y})$ are the uncorrelated Gaussian
white noises, which are used for modelling of the heat supply \cite{gavrilov2018heat,gavrilov2019steady}):
\begin{gather}
\langle
\dot {\bm W}(\breve{\bb x})
\dot {\bm W}(\breve{\bb y})^\top
\rangle=\bm I\breve\delta
(\breve{\bb x}-\breve{\bb y});
\end{gather}
$\breve\delta (\breve{\bb x})$ is a kind of the Kronecker delta with a
vectorial argument:
\begin{gather}
\breve\delta
(\breve{\bb x})\=
\left\{
\begin{aligned}
&1,&\quad &\breve{\bb x}=\bb0,\\
&0,&\quad &\breve{\bb x}\neq\bb0;
\end{aligned}
\right.
\label{gra-Kronecker}
\end{gather}
$\eta_0$ is the external viscosity. 
Note that the equations of motions in the form of 
Eqs.~ \eqref{gra-maineq}, \eqref{gra-v-dot-u}
assume the external random excitation
to be equal for all particles in a cell. 
In what follows, {we use specific quantities}:
\begin{gather}
\eta=\frac{\eta_0}m,\qquad
b=\frac{b_0}m.
\end{gather}
The initial conditions at $t=0$ are 
\begin{equation}
\bm u(\breve{\bb x})\big|_{t=0}=\bb 0,\qquad
{\bm v}(\breve{\bb x})\big|_{t=0}=\bb 0.
\label{gra-ic-stochastic}
\end{equation}

Linear equations 
\eqref{gra-maineq},
\eqref{gra-v-dot-u} are applicable only in the case when the graphene sheet under
consideration can be
treated as a taut discrete membrane. They assume that an initially plane sheet
in the natural state is pre-stressed by a uniform isotropic in-plane tensile loading.
The value of the pre-stress defines the bond stiffness $C$.
A non-zero bending stiffness of the sheet is
neglected. Note that without such a pre-stress the constitutive behaviour of a
graphene sheet must be essentially non-linear. The corresponding continuum model was 
developed in studies \cite{Sfyris2014a,Sfyris2014}. 

We expect that energy transport in the situation considered in our paper is
related mostly with out-of-plane membrane oscillation, since in the ``pure
membrane'' case in-plane membrane oscillation of order $\epsilon^2$
corresponds to out-of-plane oscillation of order $\epsilon$ (where $\epsilon$
is a formal small parameter). On the other hand, if in-plane oscillation with
a lower order of smallness co-exists with out-of-plane oscillation, then 
linear governing equations may become inapplicable. These facts, apparently,
can be demonstrated in the framework of a more general non-linear discrete model
in the same way as it was
done in the case of a one-dimensional continuum membrane (i.e.~a string) in
studies \cite{Gavrilov(ActaMech),Gavrilov2016,Ferretti2019,FerrettiJSV2019}.

\begin{remark}  
\label{gra-remark-arra}
Equations 
\eqref{gra-maineq},
\eqref{gra-v-dot-u}
are exact discrete equations of motion for the lattice under consideration, but 
as usual for a polyatomic lattice (see e.g \cite{Kuzkin2019,kosevich2005the}),
they are formulated in the form, 
which does not take into account an arrangement of the particles
inside a primitive cell.  Namely, the solution 
is defined for integer values of co-ordinates $\breve{x}^\beta$, which play a
role of generalized co-ordinates. In our case, these generalized co-ordinates
have the same value for green and red particles inside a primitive cell (see
Fig.~\ref{graphene.pdf}), which correspond to the spatial position 
$\breve x=\breve x^\beta\bb b_\beta$
of the
corresponding blue cell centre. At the same time the exact spatial positions 
for the red and green particles are $\breve x^\beta\bb b_\beta\pm (a/2)\,\bb e_2$.
This fact may be crucial for the continualization procedure,
where a vectorial continuum variable $\bb x=x^\beta \bb b_\beta$ is introduced
instead of 
$\breve{\bb x}$,
considering $\breve x^\beta$ as the {\it spatial}
co-ordinates for particles inside a cell (which is clearly not true in our
case).
In particular, the continualization procedure used in \cite{Kuzkin2019} 
(which results will be used in the present paper)
does not take into account that the spatial co-ordinates of the particles differ
from the corresponding generalized co-ordinates.
For a primitive scalar lattice (where $N=1$) situation becomes simpler,
since the generalized co-ordinates coincide with the spatial co-ordinates.
Thus, we generally expect that the results obtained for a polyatomic lattice 
in the framework of such a continuum approximation may be in a bit worse
agreement with results of
a discrete consideration,
comparing with the case of a primitive scalar lattice.
\end{remark}

Following to \cite{kuzkin2019thermal}, we introduce the temperature matrix
$\breve{\bm T}(\breve{\bb x})$
\begin{equation}
\breve{\bm T}(\breve{\bb x})\=mk_B^{-1}
\langle
{\bm v}(\breve{\bb x})
{\bm v}(\breve{\bb x})^\top
\rangle
\label{gra-Tmatrix}
\end{equation}
and the conventional kinetic temperature for a primitive cell
\begin{equation}
\breve T(\breve{\bb x})=\frac1N \tr \breve{\bm T}(\breve{\bb x}).
\label{gra-Tmatrix-trace}
\end{equation}
We also introduce the heat supply matrix $\bm B$
\begin{gather}
\bm B(\breve{\bb x},t)\=
\frac {mk_B^{-1}b^2(\breve{\bb x},t)}2
\langle
\dot {\bm W}(\breve{\bb x})
\dot {\bm W}(\breve{\bb x})^\top
\rangle
=\breve\chi(\breve{\bb x},t)\bm I,
\label{gra-heat-supply-ma}
\end{gather}
where $\breve\chi$ is the heat supply 
{per particle in a cell:}
\begin{gather}
\breve\chi(\breve{\bb x},t)
=
\frac1N \tr \breve{\bm B}(\breve{\bb x})=
\frac {mk_B^{-1}b^2(\breve{\bb x},t)}2.
\label{gra-chio}
\end{gather}
The physical meaning of the 
factor $1/2$ in 
Eqs.~\eqref{gra-heat-supply-ma}, \eqref{gra-chio}
is related to 
the fact that a half of supplied kinetic energy transforms into the potential
energy of the bonds.
This multiplier also emerges in the expressions for the heat supply for a
one-dimensional chain and a two-dimensional scalar lattice
\cite{gavrilov2019steady,gavrilov2018heat}.

The aim of the study is to find the kinetic temperature $T(\breve{\bb x},t)$
for the given
heat supply $\breve\chi(\breve{\bb x},t)$.
In what follows, we are mostly interested in the steady-state solution describing
the kinetic temperature distribution caused by a point heat source 
\eqref{gra-chio}
\begin{equation}
\breve\chi(\breve{\bb x},t)=
\breve\chi_0(t)
\breve\delta
(\breve{\bb x})
\label{gra-point-source1}
\end{equation}
{of constant intensity}
\begin{equation}
\breve\chi_0(t)=
\chio=\mathrm{const}>0.
\label{gra-point-source2}
\end{equation}

\subsection{The primitive rhombic scalar lattice of graphene cell centres}
\label{gra-sec-rhombic-statement}
In Sect.~\ref{gra-sec-rhombic-sol} we look for the solution of an 
auxiliary problem concerning the primitive rhombic scalar lattice 
(the lattice of blue circles in
Fig.~\ref{graphene.pdf}), which is 
analogous to one formulated in 
Sect.~\ref{gra-sec-formulation-gra}
for graphene lattice.
The formulation
of such a problem can be formally obtained by 
considering equations from Sect.~\ref{gra-sec-formulation-gra}, taking 
$N=1$ instead of Eq.~\eqref{gra-Nis2},
and 
\begin{equation}
       \bm C_0 = 2C,\qquad \bm C_1 = -C
\end{equation}
instead of Eq.~\eqref{gra-C-matrices}.
Here we do not distinguish $1\times1$ matrices and scalars.

\section{Approximate continuum solution}
\label{sec-gra-approximate}
\subsection{The primitive rhombic scalar lattice}
\label{gra-sec-rhombic-sol}
For the primitive rhombic scalar 2D lattice,
the kinetic temperature distribution in this case can be found
by an approach 
used in studies 
\cite{gavrilov2018heat,gavrilov2019steady,Kuzkin-Krivtsov-accepted}.

In \cite{gavrilov2019steady} a square scalar lattice is under consideration. The
steady-state solution for the kinetic temperature $T$ is obtained 
for the case of the point heat source
\eqref{gra-point-source1}
of a constant intensity
\eqref{gra-point-source2}
and a positive dissipation $\eta>0$.

In \cite{Kuzkin-Krivtsov-accepted} a general two-dimensional primitive scalar
lattice is considered. 
The expression for the kinetic temperature $T$ is obtained {for the case}
\begin{gather}
\eta=0,
\label{gra-kuzkin-eta0}
\\
\breve\chi(\breve{\bb x},t)=0.
\label{gra-chi-is-zero}
\end{gather}
Instead of non-zero heat supply $\breve\chi$, the following non-zero initial
conditions for the kinetic temperature are specified:
\begin{equation}
 \breve T\big|_{t=0}={\breve T_0}(\breve {\bb x}),\qquad \dt {\breve T}\big|_{t=0}=0.
\label{gra-kuzkin-ic}
\end{equation}
Using the It\^o lemma \cite{stepanov2013stochastic} and apparatus of 
generalized functions \cite{Vladimirov1971}
one can show (analogously to \cite{gavrilov2018heat}) that 
adopting Eqs.~\eqref{gra-chi-is-zero}, \eqref{gra-kuzkin-ic} together
is physically equivalent to specifying zero initial conditions instead of 
Eq.~\eqref{gra-kuzkin-ic}
and taking $\breve\chi(\bb x, t)$ in the following form 
\begin{equation}
\breve\chi(\breve{\bb x},t)=\frac{\breve T_0(\breve{\bb x})}2\,\delta(t).
\label{gra-Kuzkin-assumes}
\end{equation}
The initial temperature $\breve T_0(\breve{\bb x})$ is assumed to be a slowly-varying
function of the discrete vectorial variable $\breve{\bb x}$. In principle, 
this assumption allows
one to introduce naturally the initial temperature $T_0({\bb
x})$ as a continuous function of a continuum vectorial spatial variable 
\begin{equation}
\bb x=x^\beta\bb b_\beta,\qquad x^\beta\in\mathbb R,
\qquad
x^\beta=\breve x^\beta \quad \mathrm{if}\quad x^\beta\in\mathbb Z,
\label{gra-spatial-x}
\end{equation}
and to define a continuum quantity $\chi(\bb x,t)$ accordingly to
Eq.~\eqref{gra-Kuzkin-assumes}.
Note that 
the  factor $1/2$ in Eq.~\eqref{gra-Kuzkin-assumes} emerges 
following to Eqs.~\eqref{gra-Tmatrix}--\eqref{gra-chio}
and is again associated with the fact that a half of supplied kinetic energy transforms into the
potential energy of the bonds (so-called thermal equilibration
\cite{hemmer1959dynamic,klein1953mecanique,krivtsov2014energy,kuzkin2019thermal,Gavrilov2019,Sokolov2021,Berinskii2020}).
\begin{remark}  
The question about the best continuum approximation for a lattice 
solution defined only at integer
values of a spatial co-ordinate is discussed in book by Kunin~\cite{Kunin1982}.
\end{remark}


As it has been already discussed in 
Introduction (Sect.~\ref{gra-sec-intro}),
the approach used in both studies 
\cite{gavrilov2019steady,Kuzkin-Krivtsov-accepted}
is based on introducing and dealing with 
infinite set of covariance variables. Applying the It\^o lemma allows one
to derive an infinite deterministic system of ordinary differential equations
which follows from the equations of stochastic dynamics. Then 
{the procedure of
continualization is applied} to rewrite the finite difference operators
involved in the equation for covariance variables as compositions of finite difference
operators and operators of differentiation with respect to a spatial
continuum variable. 
The next step is the separation of slow motions, which are related to heat
propagation.
Following to 
\cite{gavrilov2019steady,Kuzkin-Krivtsov-accepted},
we now consider the kinetic temperature as a
slowly spatially-varying continuous function 
$T(\bb x,t)$
of a continuum vectorial spatial variable 
$\bb x$
defined by Eq.~\eqref{gra-spatial-x}.
Finally, in
\cite{Kuzkin-Krivtsov-accepted,gavrilov2019steady}
it is shown that in a continuum
approximation the kinetic temperature $T$ can be found as the inverse discrete-time
Fourier transform $\theta_F(p_1,p_2)$ of a covariance variable $\theta_{mn}$:
\begin{equation}
\kappaT\=\theta_{00},\qquad \theta_{mn}=\frac 1{(2\pi)^d}\iint_{-\pi}^\pi \theta_F(p_1,p_2)
\exp(\I mp_1+\I np_2)\,\d p_1\,\d p_2,
\label{gra-2d-maineq-prealpha0}
\end{equation}
where $\theta_F(p_1,p_2)$ is the solution of the following PDE
\begin{equation}
(\dt+\eta)^2\theta_F-
({\groopv}\cdot\nabla)^2\theta_F
=
(\dt+\eta)\chi,
\label{gra-2d-statics}
\end{equation}
which vanishes%
\footnote{Everywhere, excluding the singular rays in the case when the steady-state
solution is considered, see Sect.~\ref{gra-sec-singular}.} at infinity.
Here
$ \groopv={\groopv}(p_1,p_2) $
is the vector of group velocity for the lattice;
\begin{equation}
\nabla=
\pd{}{x^\gamma}\bb b^\gamma
\end{equation}
is the nabla operator,
$\bb b^\gamma$ such that
\begin{equation}
\bb b^1=\frac1{3a}\big(\sqrt3\,\bb e_1 +\bb e_2 \big),
\qquad
\bb b^2=\frac1{3a}\big(-\sqrt3\,\bb e_1 +\bb e_2 \big)
\label{gra-bbe-dual}
\end{equation}
is the basis dual to the basis $\bb b_\beta$:
\begin{gather}
\bb b_\beta\cdot\bb b^\gamma=\delta_\beta^\gamma,
\end{gather}
$\chi$ is the continuum approximation for $\breve\chi$ as discussed
after formula 
\eqref{gra-spatial-x}.
\begin{remark}  
Despite the fact that 
Eq.~\eqref{gra-2d-statics} was derived using variables $x^\beta$, finally the
left-hand side of this
equation is formulated in an invariant form. 
The operator in the left-hand side depends on the group velocity, which can
be, in principle, calculated in any basis. 
\end{remark}

For initial conditions 
\eqref{gra-kuzkin-ic} formula 
\eqref{gra-2d-maineq-prealpha0}
for the kinetic temperature can be written as  
follows~\cite{Kuzkin-Krivtsov-accepted}:
\begin{equation}
T=\frac{H(t)}{4(2\pi)^d}
\iint_{-\pi}^\pi
\Big(
T_0\big(\bb x-{\groopv}t\big)
+
T_0\big(\bb x+{\groopv}t\big)
\Big)
\,\d p_1\,\d p_2.
\label{gra-T-T-eta0-pre}
\end{equation}
\begin{remark}  
\label{gra-remark-1wave}
Note that formula \eqref{gra-T-T-eta0-pre} 
in the case 
\begin{equation}
\groopv(-p_1,-p_2)=-\groopv(p_1,p_2), 
\end{equation}
which is assumed everywhere in what follows, can be rewritten in the
following simpler way:
\begin{equation}
T=\frac{H(t)}{2(2\pi)^d}
\iint_{-\pi}^\pi
T_0\big(\bb x-{\groopv}t\big)
\,\d p_1\,\d p_2.
\label{gra-T-T-eta0}
\end{equation}
\end{remark}

In the dissipative case $\eta>0$ this formula becomes
\begin{equation}
T=\frac{H(t)\exp{(-\eta t)}}{2(2\pi)^d}
\iint_{-\pi}^\pi
T_0(\bb x-{\groopv}t\big)
\,\d p_1\,\d p_2.
\label{gra-T-T}
\end{equation}
The latter result can be obtained as a straightforward
generalization for results of studies 
\cite{Kuzkin-Krivtsov-accepted,gavrilov2019steady,gavrilov2018heat}.
The steady-state solution, which corresponds {to the case} 
\begin{equation}
\breve\chi(\breve{\bb x},t)={\breve \chi_0(\breve{\bb x})}\,H(t),\qquad
\eta>0,
\label{gra-Ht}
\end{equation}
can be obtained by the time integration as the limiting case of the corresponding
non-stationary solution:
\begin{equation}
T=\frac{1}{(2\pi)^d}
\int_0^\infty
\exp{(-\eta \tau)}
\iint_{-\pi}^\pi
\chi_0(\bb x-{\groopv}\tau)
\,\d p_1\,\d p_2\, \d\tau,
\label{gra-T-exp-int}
\end{equation}
where Eq.~\eqref{gra-Kuzkin-assumes} is taken into account.
Here a slowly-varying continuous function $\chi_0(\bb x)$
approximates $\breve\chi_0(\breve{\bb x})$ from Eq.~\eqref{gra-Ht} in the same
way as ${T}_0(\bb x)$ approximates $\breve{T}_0(\breve{\bb x})$.

Solutions \eqref{gra-T-T-eta0},
\eqref{gra-T-T},
\eqref{gra-T-exp-int} can be formally used in the case of a point heat source 
\eqref{gra-point-source1}. 
In \cite{gavrilov2018heat,gavrilov2019steady}
it was 
demonstrated that to do this one needs to {``approximate'' 
the Kronecker delta $\breve\delta(\breve{\bb x})$} 
\eqref{gra-Kronecker}
by the Dirac delta-function,
and use such approximations as quantities $T_0$ or $\chi_0$ in the
expressions for corresponding solutions.
In this way, formula for the steady-state kinetic temperature distribution in
a one-dimensional damped harmonic crystal was obtained in
\cite{gavrilov2018heat}.
For more complicated problems it is preferable to use
an alternative approach%
\cite{gavrilov2019steady}
and look for the stationary solution  of 
Eq.~\eqref{gra-2d-statics}. In the case of a one-dimensional
damped harmonic crystal one can easily verify
that results obtained by these two approaches are equivalent, but the second
one is essentially easier.

Using this alternative approach
in \cite{gavrilov2019steady}
the following expression for kinetic temperature
in the case of the {\it square} lattice
was obtained 
\begin{equation}
\kappaT_{\mathrm{square}}=
\frac{a^2}
{(2\pi)^d}\,
\frac \chio{2|\bb x|}
\iint_{-\pi}^{\pi}
\exp
\bigg(-\frac{\eta|\bb x\cdot\hat{{\groopv}}|}{|{\groopv}|}\bigg)
\,\delta_{(1)}^{\bb b}\big(\hat{\bb x}_\perp\cdot{{\groopv}}\big)
\,\d p_1\, \d p_2.
\label{gra-2d-maineq-prealpha}
\end{equation}
{Here and in what follows},
$\delta_{(1)}^{\bb b}$ is the one-dimensional Dirac delta-function in
the space of the covariant co-ordinates $p_1,\ p_2$ related to the dual basis $\bb b^\gamma$;
\begin{equation}
\hat {\mathbf a}\=\frac {\mathbf a}{|\mathbf a|}
\end{equation}
for any non-zero vector $\mathbf a$;
\begin{equation}
\hat{\mathbf a}_\perp\=(\mathbf e_1\times\mathbf e_2)\times \hat{\mathbf a}
\end{equation}
for any unit vector $\hat{\mathbf a}$; $\times$ is the cross product. Formula 
\eqref{gra-2d-maineq-prealpha} corresponds to the following ``approximation''
of the dimensionless Kronecker delta by the dimensionless Dirac delta, which is used 
in \eqref{gra-point-source1} 
and in the right-hand side of Eq.~\eqref{gra-2d-statics}:
\begin{gather}
\breve\delta(\breve{\bb x})\approx \delta_{\bb b}(\bb x)\=\delta(x^1,x^2).
\label{gra-dirac-delta}
\end{gather}
The multiplier $a^2$ is dropped in our previous paper
(see\cite{gavrilov2019steady}, formula (5.24)). 
This is because in \cite{gavrilov2019steady}
the corresponding formula
is a kind of fundamental solution, which expresses the continuum
solution caused by the source 
\begin{gather}
\delta_{\bb e}(\bb x)\=\delta(X^1,X^2).
\end{gather}
Here ${X^\beta}$ are co-ordinates,
which correspond to the orthonormal basis $\bb e_1$, $\bb e_2$:
\begin{gather}
\bb x={X^\beta}\bb e_\beta.
\end{gather}
Note that in the case of the square lattice one has:
\begin{equation}
\delta_{\bb b}(\bb x)=a^2\,\delta_{\bb e}(\bb x).
\end{equation}

Now we want to generalize formula \eqref{gra-2d-maineq-prealpha} to the 
case of the rhombic lattice.
It is clear that the exact discrete solution for 
the rhombic lattice and the one for the square lattice are identically equal:
\begin{equation}
\breve T_{\mathrm{rhombic}}(\breve{x}^\beta)
\equiv
\breve T_{\mathrm{square}}(\breve{x}^\beta)
\end{equation}
(these two lattices are described by the same dynamical equations, and only
spatial arrangement of particles is different). Thus, the continuum solution to be found
should coincide with solution \eqref{gra-2d-maineq-prealpha}, which is
expressed in the dimensionless co-ordinates $x^\beta$.
On the other hand, in what follows, we prefer to work with co-ordinates $X^\beta$.
Now, the easiest way to
proceed with the calculations in the case of the rhombic lattice
is to return back to Eq.~\eqref{gra-2d-statics}, wherein we should again use
``approximation'' \eqref{gra-dirac-delta} in the right-hand side.
Then we rewrite
Dirac delta-function \eqref{gra-dirac-delta} in the right-hand side of
Eq.~\eqref{gra-2d-statics} using co-ordinates ${X^\beta}$.
One has \cite{Vladimirov1971}
\begin{gather}
\delta_{\bb b}(\bb x)=|\Upsilon|\,\delta_{\bb e}(\bb x),
\end{gather}
where $|\Upsilon|$ is the absolute value of the determinant for the {matrix of covariant
transformation}~\cite{ZhilinVectors}:
\begin{equation}
\Upsilon
=
\det
\begin{pmatrix}
\bb b_1\cdot \bb e_1 & \bb b_1\cdot \bb e_2 \\
\bb b_2\cdot \bb e_1 & \bb b_2\cdot \bb e_2 
\end{pmatrix}
=a^2\det
\begin{pmatrix}
\frac{\sqrt3}2 & \frac32 \\
-\frac{\sqrt3}2 & \frac32
\end{pmatrix}
=\frac{3\sqrt3a^2}{2}.
\end{equation}
Since the left-hand side of Eq.~\eqref{gra-2d-statics} is formulated in the
invariant form, we can straightforwardly repeat all the calculations 
from \cite{gavrilov2019steady}
using the orthonormal co-ordinates
$X^\beta$. This yields
\begin{equation}
T_{\mathrm{rhombic}}(\bb x)
=\frac{|\Upsilon|}{a^2}\,
T_{\mathrm{square}}(\bb x)
=\frac 
{|\Upsilon|}
{(2\pi)^d}\,
\frac \chio{2|\bb x|}
\iint_{-\pi}^{\pi}
\exp
\bigg(-\frac{\eta|\bb x\cdot\hat{{\groopv}}|}{|{\groopv}|}\bigg)
\,\delta_{(1)}^{\bb b}\big(\hat{\bb x}_\perp\cdot{{\groopv}}\big)
\,\d p_1\, \d p_2.
\label{gra-2d-maineq-alpha}
\end{equation}
where 
$T_{\mathrm{square}}(\bb x)$ is defined by Eq.~\eqref{gra-2d-maineq-prealpha}.
We have verified the last formula by numerical calculations using the same
approach with that we use for a square lattice in
\cite{gavrilov2019steady}, and an excellent agreement has been obtained.

\subsection{The graphene lattice}
\label{gra-sect-gra}
The non-stationary propagation of the kinetic temperature field in an undamped 
($\eta=0$)
polyatomic lattice of general structure described by 
Eqs.~\eqref{gra-maineq}--\eqref{gra-v-dot-u} was considered in recent paper by Kuzkin~\cite{Kuzkin2019}, where the
expression for the kinetic temperature is obtained, in particular, for the
case\footnote{Kuzkin actually
considered
a more general problem formulation:
the external random excitation is not assumed to be equal for all particles
in a cell
(Eq.~\eqref{gra-heat-supply-ma} generally is not assumed to be true),
and the masses of the particles inside a cell are also not assumed to be
equal.}
\eqref{gra-Kuzkin-assumes}, which is 
equivalent to choosing of {zero heat supply}
\eqref{gra-chi-is-zero}
{and the following non-zero initial conditions for the temperature matrix:}
\begin{equation}
\breve{\bm T}\big|_{t=0}={T_0}\bm I,\qquad \dt \breve{\bm
T}\big|_{t=0}=\mathbf 0,
\end{equation}
which are analogous to \eqref{gra-kuzkin-ic}.

{As it has been already discussed in Introduction (Sect.~\ref{gra-sec-intro}),
Kuzkin in \cite{Kuzkin2019}
uses a completely different from \cite{gavrilov2019steady}
technique to solve the problem and evaluate a continuum approximation 
$T(\bb x, t)$ for $\breve T(\breve {\bb x}, t)$. At the
first step, the exact expression in the form of a multiple integral
for the temperature matrix $\breve{\bm T}(\breve {\bb x}, t)$
is found. At the second step, several integrals are evaluated using an approximate
approach which is related,  apparently, with the method of stationary phase.
Finally, the kinetic temperature is found as the sum of the slow component and
the fast one.} In the case of a 2D polyatomic
lattice ($d=2$), the final formula for the slow component, which is related to heat
propagation, has the following form:%
\footnote{This is the last formula in Eq.~(30) \cite{Kuzkin2019} (with
the simplification discussed in Remark~\ref{gra-remark-1wave} of the present
paper).}
\begin{equation}
T=\frac{H(t)}{2(2\pi)^dN}
\sum_{i=1}^N
\iint_{-\pi}^\pi
T_0(\bb x-{\groopv}_it)
\,\d p_1\,\d p_2,
\label{gra-T-T-N}
\end{equation}
where ${\groopv}_i(p_1,p_2)\ (i=\overline{1,N})$ are the group velocities
corresponding to all
dispersion branches for the polyatomic lattice under consideration.
Formula 
\eqref{gra-T-T-N}
generalizes 
Eq.~\eqref{gra-T-T-eta0}
to the case of a polyatomic lattice.
In this paper we will use the following formula describing the 
kinetic temperature in a polyatomic lattice:
\begin{gather}
\kappaT
=
\frac {|\Upsilon|}{(2\pi)^dN}\,\sum_{i=1}^N
\frac \chio{2|\bb x|}
\iint_{-\pi}^{\pi}
\exp
\bigg(-\frac{\eta|\bb x\cdot \hat{{\groopv}}_i|}{|{\groopv}_i|}\bigg)
\,\delta_{(1)}^{\bb b}\big(\hat{\bb x}_\perp\cdot{{\groopv}_i}\big)
\,\d p_1\, \d p_2.
\label{gra-maineq-prealpha}
\end{gather}
Formula 
\eqref{gra-maineq-prealpha}
is an informal generalization of previously obtained results (in
particular, Eq.~\eqref{gra-2d-maineq-alpha})
for a polyatomic lattice in the dissipative case. We argue the possible
applicability of formula 
\eqref{gra-maineq-prealpha}
by similarity of equation structures in pairs 
\eqref{gra-T-T-eta0} \& 
\eqref{gra-2d-maineq-alpha}
and 
\eqref{gra-T-T-N} \&
\eqref{gra-maineq-prealpha}, respectively, taking into account that 
Eqs.~\eqref{gra-T-T-eta0}, 
\eqref{gra-2d-maineq-alpha}
\eqref{gra-T-T-N} are developed in an accurate way and verified by numerical
calculations.
The accurate derivation of Eq.~\eqref{gra-maineq-prealpha}
is beyond the scope of this paper. This should be done
in a manner similar to one used in 
\cite{Kuzkin2019}. There most essential difficulty in comparison with the
procedure in 
\cite{Kuzkin2019} is the presence of dissipation ($\eta>0$). 
%

\section{Simplifying the continuum solution}
\label{gra-sec-simplify}

In this section we proceed with the simplification of 
formula~\eqref{gra-maineq-prealpha} in the case of graphene lattice and
provide the corresponding calculations.

Since the integrand in the right-hand side of Eq.~\eqref{gra-maineq-prealpha}
involves one-dimensional Dirac delta $\delta_{(1)}^{\bb b}$, the corresponding double integral can be reduced to
a single one. The technique is analogous to one used in 
\cite{gavrilov2019steady}.
Namely, we need to use
the formula for the composition of the Dirac delta and a smooth function 
(see \cite{G-Sh-1})
\begin{equation}
\delta_{(1)}\big(f(\qy)\big)\,\d\qy=\sum_j \frac{\delta_{(1)}(\qy-\qy_j)}{|f'(\qy_j)|},
\label{gra-int-delta}
\end{equation}
where $y_j$ are the roots of $f (y)$. 

\begin{figure}[p]	
\centering\includegraphics[width=\textwidth]{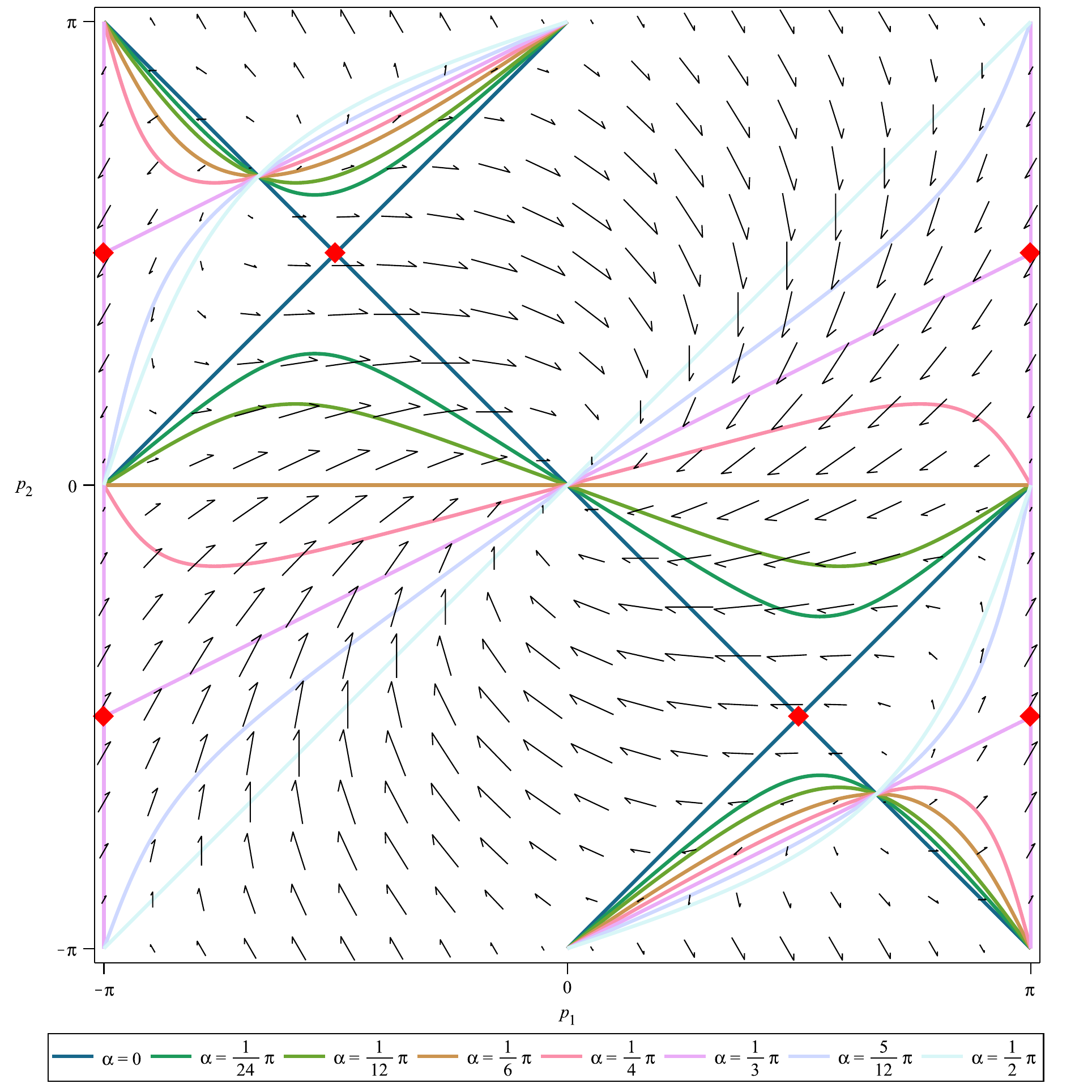}%
\caption{The vector field ${\Groopv}(p_1,p_2)$ characterizing the
directions of both group velocities $\bC_\pm(p_1,p_2)$ and family of curves defining the
roots $p_2^{(j)}(p_1,\alpha)$. Along the every curve the directions of the
group velocities $\bC_\pm$ are fixed, in such a way that Eq.~\eqref{gra-proots} is
fulfilled. The red diamonds correspond to points,
which bring the singular contributions into the analytical continuum solution
(these are the intersections of curves defining the different roots $p_2^{(j)}(p_1,\alpha)$
for $\alpha\in\{0;\frac\pi3\})$
}
\label{gra-p-roots.pdf}
\end{figure}

\begin{figure}[htbp]	
\centering\includegraphics[scale=0.5]{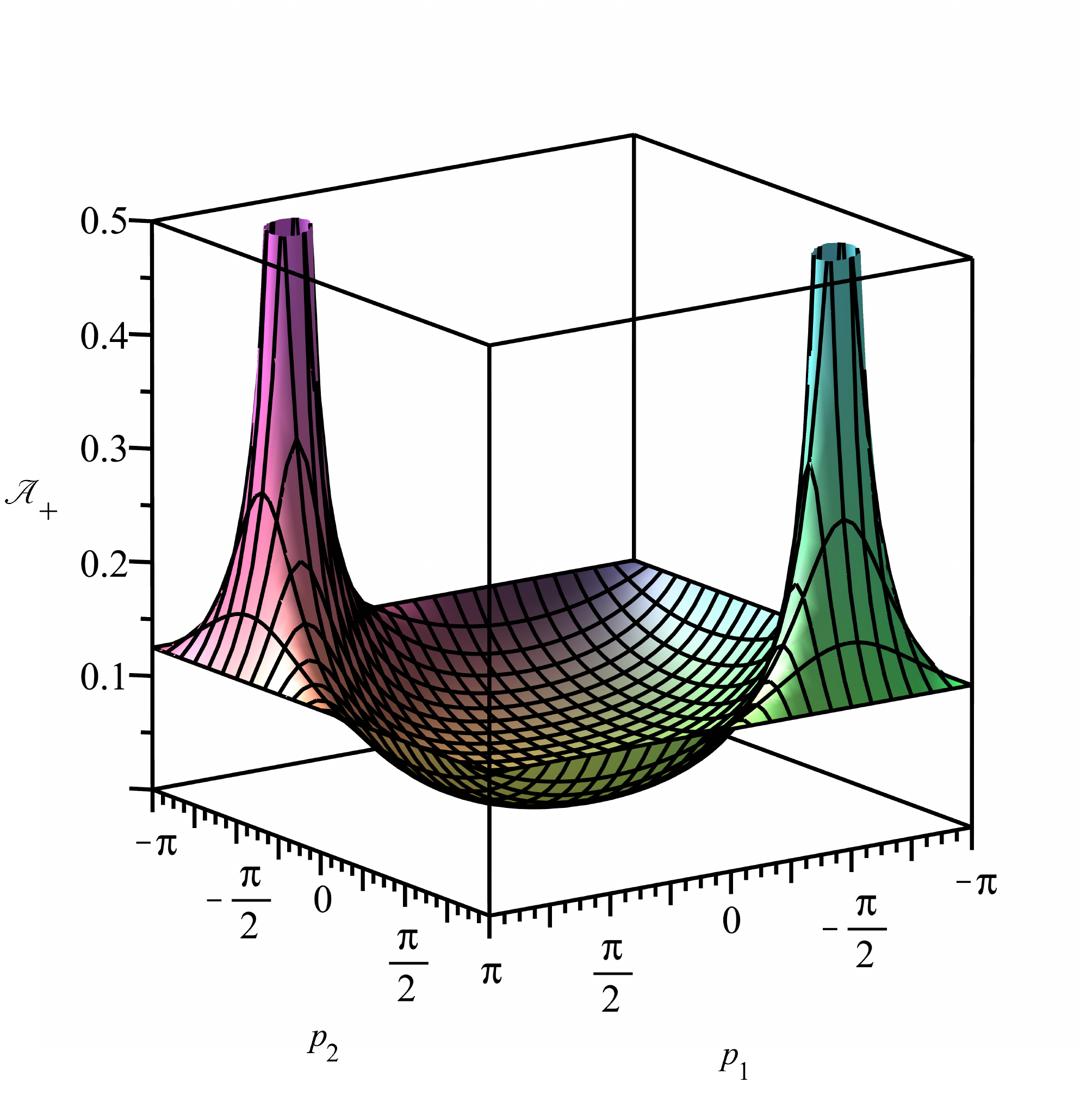}\\
\centering\includegraphics[scale=0.5]{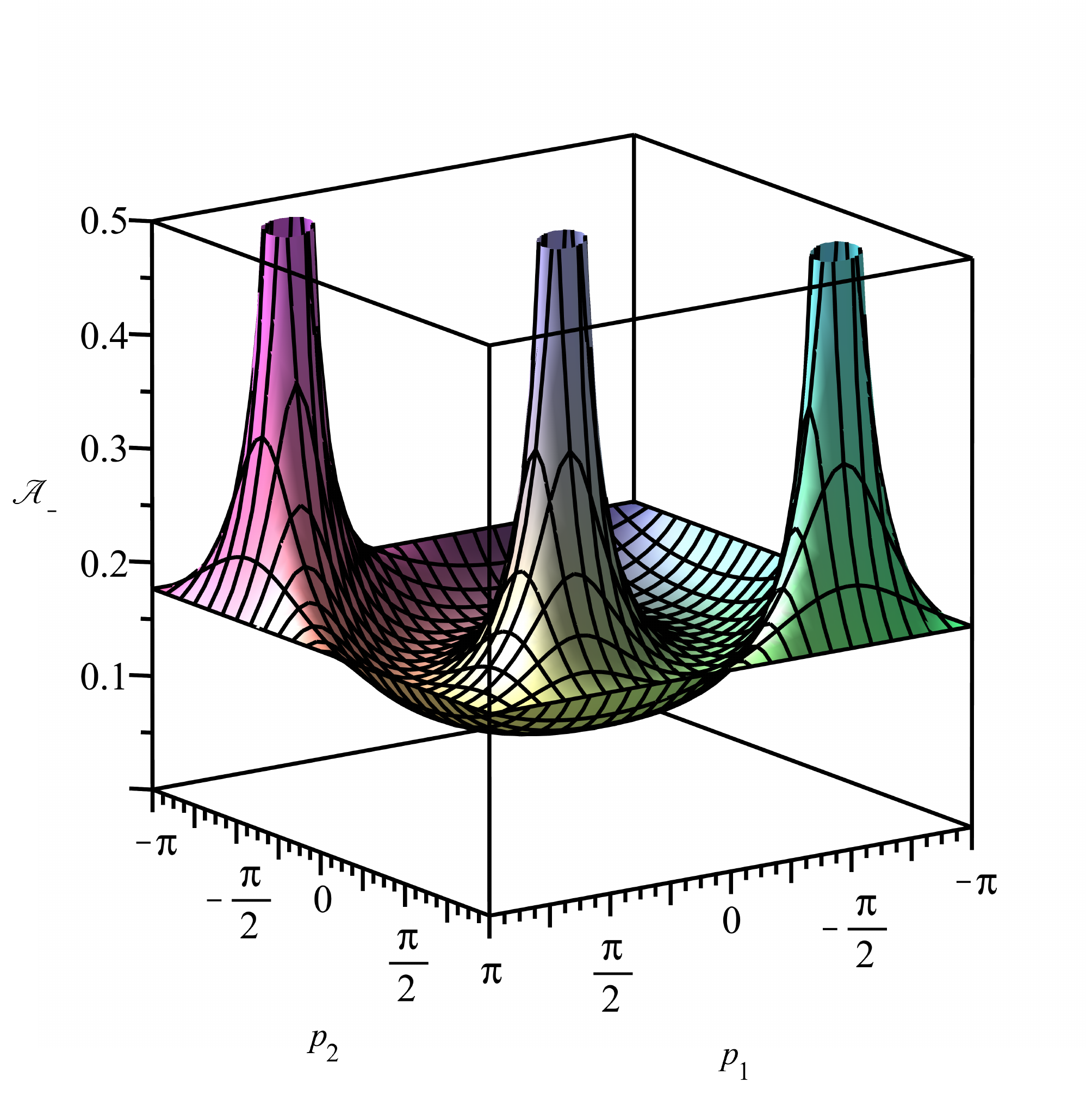}
\caption{3D plots of functions $\mathscr A_\pm(p_1,p_2)$ ($c=1$)}
\label{gra-A-plot}
\end{figure}

Since a graphene lattice has two planes of symmetry, which are orthogonal to
vectors $\bb e_1$ and $\bb e_2$, without loss of generality we can assume that
$0\leq\alpha<\pi/2$, where
\begin{gather}
\alpha=\arctan \frac {{X^2}}{{X^1}}.
\end{gather}
The expressions for the dispersion surfaces and the group velocities 
for graphene lattice 
\cite{kuzkin2019thermal,Kuzkin2019,Abdukadirov_2019,AyzenbergStepanenko2008}
can be found in Appendix~\ref{gra-sec-app-dispersion}.
In order to simplify formula \eqref{gra-maineq-prealpha},
it is useful to represent the vector fields of group velocities
$\bC_\pm\equiv\bC_{1,2}$ \eqref{gra-group-initial} in the following form
\begin{gather}
\bC_\pm(p_1,p_2)=\mp \mathscr A_\pm(p_1,p_2) {\Groopv}(p_1,p_2),
\\
\mathscr A_\pm(p_1,p_2)=\frac{\sqrt3c}{R(p_1,p_2)\sqrt{3\pm R(p_1,p_2)}},
\\
{\Groopv}(p_1,p_2)=\big(2\sin(p_1-p_2)+\sin p_1-\sin p_2\big)\bb e_1
+\sqrt3(\sin p_1+\sin p_2)\bb e_2.
\end{gather}
Here $R(p_1,p_2)$ is defined by 
Eq.~\eqref{gra-R-p1p2}, and $c$ is defined by 
\eqref{gra-c-def}.
The vector field ${\Groopv}(p_1,p_2)$ characterizes the directions of
both group velocities $\bC_\pm(p_1,p_2)$; it
is shown in Fig.~\ref{gra-p-roots.pdf}.
3D plots for quantities ${\mathscr A}_\pm(p_1,p_2)$ are shown in
Fig.~\ref{gra-A-plot}. One has 
\begin{equation}
\mathscr A_\pm>0
\end{equation}
for $-\pi\leq p_1\leq\pi$, $-\pi\leq p_2\leq\pi$. Quantities ${\mathscr
A}_\pm(p_1,p_2)$ are singular 
\begin{equation}
\mathscr A_\pm\to+\infty
\end{equation}
at the conical points $p_1=2\pi/3=-p_2$ and $p_1=-2\pi/3=-p_2$, where the
dispersion surfaces come into the contact (see
Fig.~\ref{gra-omega-plot}, Appendix~\ref{gra-sec-app-dispersion}). In
neighbourhoods of the conical points the absolute values $|\bC_\pm|$ of both 
group velocities
$\bC_\pm$ are finite quantities, whereas their directions quickly change following to the
direction of ${\Groopv}$.
Additionally,
$\mathscr A_-\to+\infty$ at $p_1=p_2=0$, where the group velocity $\bC_-$ is
zero.
\begin{figure}[p]	
\centering\includegraphics[width=\textwidth]{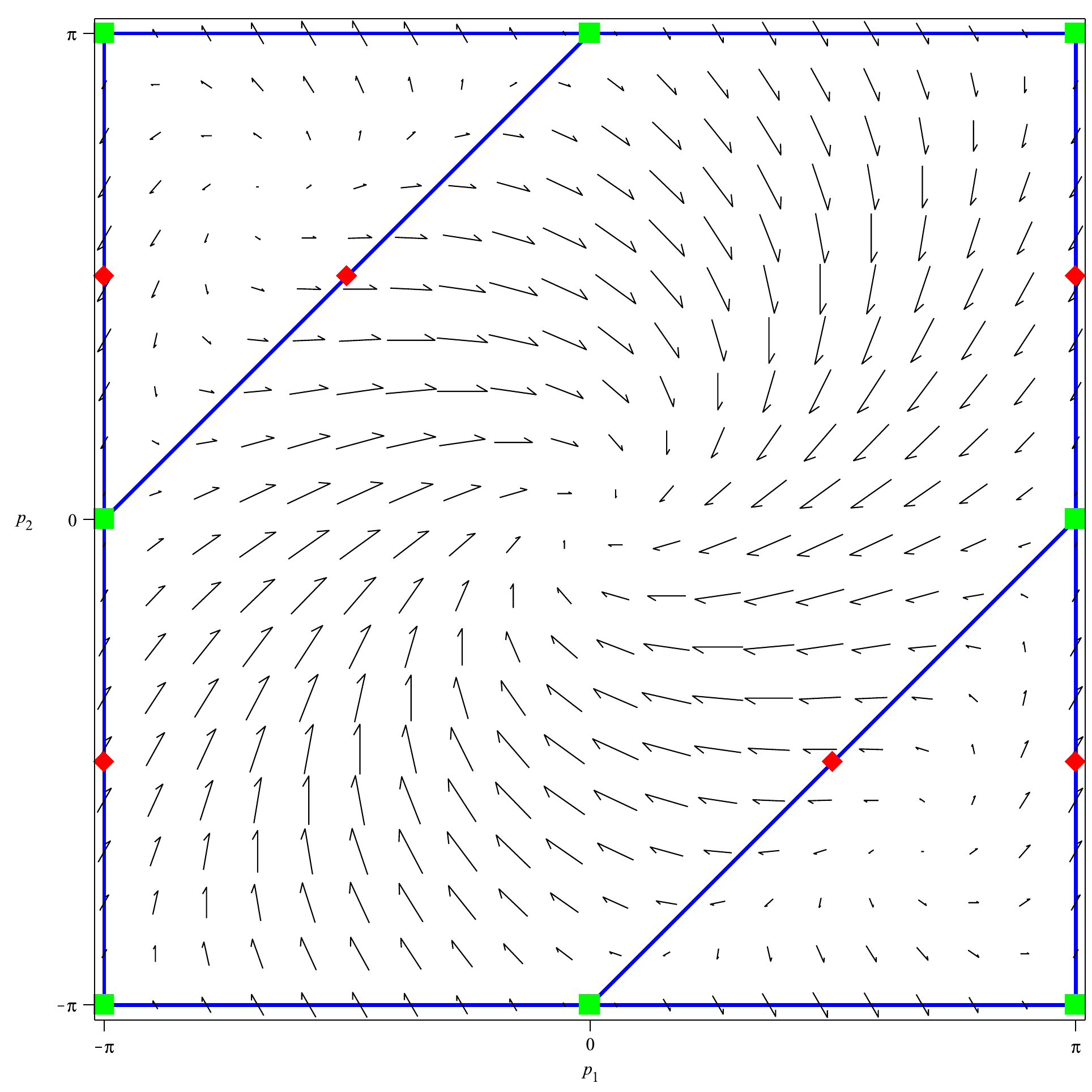}
\caption{The vector field ${\Groopv}(p_1,p_2)$ characterizing the
directions of both group velocities $\bC_\pm(p_1,p_2)$ and the projections on
plane $\omega=0$ of the curves defining the contribution from resonant
frequencies $\omega_\pm(p_1,p_2)=\bar\omega_\pm$ (the blue lines). 
The red diamonds correspond to points,
which bring the singular contributions into the analytical continuum solution
(these are the intersections of curves defining the different roots $p_2^{(j)}(p_1,\alpha)$
for $\alpha\in\{0;\frac\pi3\}$, see Fig.~\ref{gra-p-roots.pdf}). The green
boxes are points where both group velocities are zero: 
$\bC_\pm(p_1,p_2)={\Groopv}(p_1,p_2)=\mathbf 0$}
\label{gra-resonance.pdf}
\end{figure}

One has
\begin{gather}
\hat\bC_\pm=\mp
\frac{{\Groopv}}{|{\Groopv}|},
\\
\strut
|{\Groopv}|=\sqrt{\big(2\sin(p_1-p_2)+\sin p_1-\sin p_2\big)^2+3(\sin p_1+\sin
p_2)^2},
\\
\bb x_\perp=|\bb x|(\cos \alpha\, \bb e_2- \sin \alpha\, \bb e_1),\\
\hat{\bb x}_\perp\cdot\bC_\pm
=\breveC_\pm(p_1,p_2)\Psi(p_1,p_2),
\\
\Psi(p_1,p_2)=\bb x_\perp\cdot{\Groopv},
\\
\pd{\Psi}{p_2}
=
\cos(p_1-p_2)\sin\alpha
+\frac{\cos p_2\sin\alpha}2
+
\frac{\sqrt3\cos p_2\cos\alpha}2
.
\label{gra-dPsi}
\end{gather}
Thus,
\begin{gather}
\hat{\bb x}_\perp\cdot\bC_\pm=0
\quad\Longleftrightarrow\quad
\Psi=0
\quad
\Longleftrightarrow\quad
p_2=p_2^{(j)}(p_1)\in[-\pi,\pi],\quad j=1,2.
\label{gra-proots}
\end{gather}
{The plot of roots $p_2^{(j)}$ for various values of $\alpha$
is demonstrated in Fig.~\ref{gra-p-roots.pdf}.} Note that in the particular
case $\alpha=\pi/3$ one need to represent the roots in the alternative form 
$p_1=p_1^{(j)}(p_2)$ (see Fig.~\ref{gra-p-roots.pdf}); the subsequent
analysis is analogous to the case $\alpha\neq\pi/3$.
For an arbitrary $0\leq\alpha\leq\pi/2$ the expressions for roots $p_2^{j}$
are very complicated and lengthy, therefore we have used {\sc Maple} symbolic
calculation software to obtain them.

Now we are ready to apply formula
\eqref{gra-int-delta} to Eq.~\eqref{gra-maineq-prealpha}. This yields
\begin{gather}
\delta_{(1)}^{\bb b}(
\hat{\bb x}_\perp\cdot\bC_\pm
)
=
\sum_{j=1}^2
\frac{\delta_{(1)}^{\bb b}\big(p_2-p_2^{(j)}(p_1)\big)}
{\breveC_\pm\left|
\pd{\Psi_\pm}{p_2}
\right|}
,\\
\kappaT=
\frac {\chio|\Upsilon|}{(2\pi)^dN|\bb x|}
\sum_{(\pm)}\sum_{j=1}^2
\bigintss_{0}^{\pi}
\left.\frac{\exp
\left(-\frac{\eta\,|\bb x\cdot\hat{{\groopv}}|}
{|\groopv_\pm|}\right)
\,\d p_1}{\breveC_\pm
\left|
\pd{\Psi_\pm}{p_2}
\right|
}
\,\right|_{p_2=p_2^{(j)}(p_1)}
.
\label{gra-2dkappa-final-app}
\end{gather}
Here we have already taken into account that 
the integrands in the right-hand side of 
Eq.~\eqref{gra-2dkappa-final-app}
turn out to be 
even functions of $p_1$.
Thus, we have got the final formula in the form of a single integral, which we
use to calculate the continuum solution
(see Sect.~\ref{gra-sec-numerics}).

\section{Singular rays and phonon focusing}
\label{gra-sec-singular}
{The numerical calculations show that formula 
\eqref{gra-2dkappa-final-app}
predicts that the steady-state continuum solution
possesses the rotational symmetry of order six (the rotation by
angle $\pi n/3$, $n\in\mathbb Z$ does not change the continuum solution).} 
Along six rays, where 
\begin{equation}
\alpha\in 
\left\{-\frac\pi3,0,\frac\pi3\right\} 
\label{gra-singular-rays}
\end{equation}
the continuum solution is singular\footnote{Note that in the case of a primitive
rhombic scalar lattice (in particular for a square lattice) the singular
rays also exist and correspond to $\alpha
\in\left\{-\frac\pi2,0,\frac\pi2\right\} 
=0$, see
\cite{gavrilov2019steady}.}. These
singular rays correspond to the existence (for such values of $\alpha$) of
multiple (double) roots of equation $\Psi=0$. If a multiple root exists, then
the derivative \eqref{gra-dPsi} becomes zero simultaneously with function~$\Psi$. This
leads to the emergence of a non-integrable singularity in the denominator of the
integrand in the right-hand side of Eq.~\eqref{gra-2dkappa-final-app}. In
Fig.~\ref{gra-p-roots.pdf}
the red diamonds correspond to points, which bring the singular contributions
into the analytical continuum solution. 
Note that (see Sect.~\ref{gra-sec-numerics} and, in particular, 
Fig.~\ref{graphene-2d.pdf}(a) \& Fig.~\ref{graphene-time.pdf}) the discrete solution
does not possess this property and remains to be finite everywhere being
localized along singular rays, see a star-like structure in the centre
of the plot in Figure~\ref{graphene-2d.pdf}(a).

In the conservative non-stochastic case the singular rays in a square
scalar lattice have been observed in
\cite{mielke2006macroscopic,harris2008energy,giannoulis2006continuum}, where
they are associated with contributions from points on dispersion surfaces,
where Jacobian 
\begin{equation}
\det \pd{(g^1,g^2)}{(p_1,p_2)}= 0.
\label{gra-Jacobian}
\end{equation}
The last relationship is clearly fulfilled at points marked 
by red diamonds in Figs.~\ref{gra-p-roots.pdf}, \ref{gra-resonance.pdf}.
Here $g^1,\ g^2$ are co-ordinates of the group velocity $\bC$ in the
lattice basis ($\bb b_\gamma$ in the case of graphene).
In physical literature
\cite{Fu2020,Northrop1980,Northrop1979,Wolfe1998}
the phenomenon of localization of the ballistic phonons along the singular
rays, which can be observed in experiments with anisotropic 3D crystals,
is known as the phonon focusing.
The phonon focusing also is associated with points on the dispersion surfaces,
where the corresponding Jacobian is zero (see \cite{Maris1971}, where the
corresponding result is obtained basing on conceptions of stationary
anisotropic continuum elastodynamics).
\begin{remark} 
Graphene layer simulated as a two-dimensional elastic continuum has
isotropic elastic properties. For in-plane motions this is proved in 
\cite{Berinskii2013}, for out-of-plane motions (an elastic membrane
\cite{Scuracchio2014}) this fact is also true and can
be proved in an analogous to \cite{Berinskii2013} way.
\end{remark}

More detailed explanation
of singular rays emergence in lattices, which is based on non-stationary
conceptions, is given in
\cite{slepyan1987energy,Abdukadirov_2019,AyzenbergStepanenko2008}.
According to these studies
the emergence of singular rays is associated with the existence {of
special localized types of solutions of deterministic equations of motions}: so-called
star waves and line-localized primitive 
waveforms. It was shown in
\cite{slepyan1987energy} that the frequencies, to which zero group velocity
corresponds, are resonant frequencies of an infinite (continuum or discrete)
mechanical system. For the graphene lattice under consideration there are at
least two such frequencies: 
\begin{equation}
\bar\omega_-=\sqrt2\omega_\ast, \qquad
\bar\omega_+=2\omega_\ast
\label{gra-resonant}
\end{equation}
for the acoustic branch and for the optic one, repsectively (see
Appendix~\ref{gra-sec-app-dispersion}).
The cross-sections of the dispersion
surfaces (see Fig.~\ref{gra-omega-plot})
with the planes $\omega=\bar\omega_\pm$ are shown in 
Fig.~\ref{gra-resonance.pdf} as the blue lines. The green
boxes are points of these lines where both group velocities are 
zero: $\bC_\pm(p_1,p_2)={\Groopv}(p_1,p_2)=\mathbf 0$. Again, as well as
in
Fig.~\ref{gra-p-roots.pdf}, 
the red diamonds correspond to points, which bring the singular contributions
into the analytical continuum solution. One can see that such points belong to
the blue lines, i.e.\ they bring the particular contributions to the solution 
from the resonant frequencies $\bar\omega_\pm$. Moreover, additional calculations show
that for points lying at the blue lines\footnote{Excluding the points, where
$\bC_\pm=\bb 0$ and the corresponding direction is undefined,
which are shown as the green boxes.},
the direction of the group velocities 
has a value, which equals to $\pi n/3$:
\begin{equation}
\arctan\frac{\bC_\pm\cdot\bb e_2}{\bC_\pm\cdot\bb
e_1}\in
\left\{-\frac\pi3,0,\frac\pi3\right\} 
\end{equation}
(see the vector field in Fig.~\ref{gra-resonance.pdf}).
The last fact means that all energy contributed from the
resonant frequencies $\bar\omega_\pm$ propagates along singular rays
\eqref{gra-singular-rays}
and form the localized solutions.

\begin{remark}  
Note that additionally to $\bar\omega_\pm$
there is the frequency $\sqrt6\,\omega_\ast$, 
to which zero group velocity ${\groopv}_+$
corresponds. The corresponding wave vector is zero, thus this frequency does
not bring any effects related to directional localization.
\end{remark}

The question appears: is it possible to fix the continuum solution 
in such a way that it would predict a bounded value for the kinetic
temperature observable in the framework of the discrete model? In our opinion,
this should be done by constructing uniform asymptotics 
\cite{Fedoruk-Saddle,temme2014} in the framework of the 
method of stationary phase, which can be used
\cite{Gavrilov2020:2006.08197}
to obtain the continuum solution.

\section{Comparison between the discrete solution and the continuum one}
\label{gra-sec-numerics}
In this section, we compare non-stationary numerical solution of the system of
stochastic ODE
\eqref{gra-maineq},\eqref{gra-v-dot-u}
calculated for large enough time 
on the one hand, 
with the steady-state  analytical continuum solution in the integral form 
\eqref{gra-2dkappa-final-app} on the other hand.
Our methodology is similar to one used in
\cite{gavrilov2019steady} for the case of a square scalar lattice.

Without loss of generality, {we can put}
\cite{gavrilov2019steady,gavrilov2018heat}:
\begin{equation}
\omega_\ast=1,\qquad a=1, \qquad m=1, \qquad k_B=1, \qquad b=1,
\label{gra-dimensionless}
\end{equation}
and deal, in what follows, with dimensionless quantities.
We consider a lattice of $(2n+1)^2$ cells ($N(2n+1)^2$ particles) with the
periodic boundary conditions.
Actually, the specific form of this boundary conditions 
is not very important in our calculations, since
we take large enough $n$ such that the quasi-waves reflections from 
the boundaries do not occur\footnote{The non-stationary solution is almost
vanishes outside the circle with radius $g_{\mathrm{max}}t$ with centre at the point
source location, where $g_{\mathrm{max}}\simeq0.897\omega_\ast a$  \cite{Kuzkin2019} is the
maximum magnitude for the vectors~$\bC_\pm$.}.
To obtain a numerical solution we use the scheme 
\begin{multline}        
\Delta{\bm v}(\breve{\bb x},t^k)
\\=
\left(
-\eta {\bm v}(\breve{\bb x},t^k)
-\bm C_0 \bm u(\breve{\bb x},t^k)
-\sum_{\beta=1}^{2}
\left(
\bm C_1 \bm u(\breve{\bb x}+\bb b_\beta,t^k)
+\bm C_1^\top \bm u(\breve{\bb x}-\bb b_\beta,t^k)
\right)
\right)
\Delta t
\\+\breve\delta(\breve {\bb x})\bm N^k
\sqrt{\Delta t},
\label{gra-scheme}
\end{multline}
\begin{gather}
\Delta{\bm u}(\breve{\bb x},t^k)=
\bm v(\breve{\bb x},t^{k+1})\Delta t,
\label{gra-v-dot-u-scheme}
\\
\bm v(\breve{\bb x},t^{k+1})=\bm v(\breve{\bb x},t^{k})+\Delta\bm v(\breve{\bb x},t^{k}),
\\
\bm u(\breve{\bb x},t^{k+1})=\bm u(\breve{\bb x},t^{k})+\Delta\bm u(\breve{\bb
x},t^{k}).
\end{gather}
Here 
$\bm N^k$
\begin{equation}
{\bm N}^k=
\begin{pmatrix}	
{N_1^k}& {N_2^k}
\end{pmatrix}^\top
\end{equation}
are columns of generated normal random numbers such that 
\begin{equation}
\langle N_\beta^k\rangle=0,\qquad
\langle N_\beta^k N_\alpha^j\rangle=\delta_{jk}\delta_{\alpha\beta}.
\end{equation}
We perform a series of $r=1\dots R$ realizations of these calculations (with
various independent 
$\bm N^k$) and get the corresponding particle velocities
$\bm v_{(r)}(\breve{\bb x},t^k)$. In
accordance with Eqs.~\eqref{gra-Tmatrix}, 
\eqref{gra-Tmatrix-trace} 
{in order to
obtain the dimensionless kinetic temperature 
we use the following formula:
\begin{equation}
\breve T(t^k)
=\frac{1}{RN}\tr\sum_{r=1}^{R} \bm v_{(r)}(\breve{\bb x},t^k) \bm v_{(r)}^\top(\breve{\bb x},t^k).
\label{gra-kin-temp-num}
\end{equation}

Numerical results  
\eqref{gra-kin-temp-num}
for the kinetic temperature
can be compared with the analytical 
steady-state solution
\eqref{gra-2dkappa-final-app},
where \eqref{gra-dimensionless} are taken into account.
Note that according to 
Eqs.~\eqref{gra-chio},
\eqref{gra-dimensionless} 
dimensionless intensity of heat supply $\chio$ in 
\eqref{gra-2dkappa-final-app} should be taken as 
\begin{equation}
\chio=\frac12.
\end{equation}
The expressions for the integrands in the right-hand side of
Eq.~\eqref{gra-2dkappa-final-app},
are very complicated and lengthy, therefore we use {\sc
Maple} symbolic calculation software to generate {\sc Cython} \cite{CYTHON} code for
the integrands. To calculate the continuum solution in the form of a single
integral, as well as to find the discrete kinetic temperature $\breve T$,
we use {\sc SciPy} software. 


The comparison between the analytical continuum solution
and the numerical discrete one is presented in
Figs.~\ref{graphene-2d.pdf}--\ref{graphene-1d.pdf}.
All calculations have been performed for the following values of the problem
parameters: $\eta=0.1$, $n=100$,
$t=150$, $R=1600$.
The time step is $\Delta{ t}=0.025$.
The continuum solution $T(\bb x)$ is evaluated over the grid $\bb x=\breve {\bb
x}$.
We use {\sc Python} library
{\sc Matplotlib} for visualization.

In Figs.~\ref{graphene-2d.pdf} \& \ref{graphene-3d.pdf} 
one can see 2D \& 3D  plots, respectively, of a central zone of the kinetic temperature
distribution pattern in the lattice. There are three subplots on each figure;
namely,
(a) the discrete numerical solution $\breve T$, (b) the continuum solution
$T$, and (c)
the ratio $T/\breve T$ are shown in the logarithmic scale. 
The singular rays \eqref{gra-singular-rays}
are shown by the white colour for subplots (b) \& (c).
The dark yellow circle region in Fig.~\ref{graphene-2d.pdf} corresponds to a
zone, where non-stationary discrete solution can be considered as a stationary
one.  Figures \ref{graphene-2d.pdf}(c) \& \ref{graphene-3d.pdf}(c) show that
the discrete solution and the continuum solution are in a good agreement everywhere
excepting the cells on and nearby the singular rays 
(where the phonon focusing takes place) and cells nearby to the heat source.

In Fig.~\ref{graphene-1d.pdf} we compare, 
(a) in the linear scale and (b) in the logarithmic one, the continuum solution 
for the row of cells $\breve x^1=\breve x^2=X^2/3$ along the armchair
direction, and the corresponding 
discrete solution. 

Comparing the results for graphene with the ones previously obtained for the
primitive square scalar lattice \cite{gavrilov2019steady}, one can observe
that the agreement between the discrete solution and the continuum one in a
zone nearby the heat source is much better in the latter case. In our opinion,
this fact is caused by the assumption of
simplified arrangement of particles in a primitive cell, which is adopted
during the continualization in the case of a polyatomic lattice
(see Remark~\ref{gra-remark-arra}).
The influence of such a simplification becomes more important in the
domains, where a large gradient of the continuum kinetic temperature is observed. 

%

Finally, in Fig.~\ref{graphene-time.pdf} we demonstrate that the discrete
solution converges to finite values at cells located on singular rays
(see, e.g., the plot for cell $\breve x^1=0,\ \breve x^2=1$),
and, in particular, at the cell where the heat source is applied ($\breve
x^1=0,\ \breve x^2=0$). This fact has been discussed previously in
Sect.~\ref{gra-sect-gra}.

\begin{figure}[p]	
\centering\includegraphics[width=0.6\columnwidth]{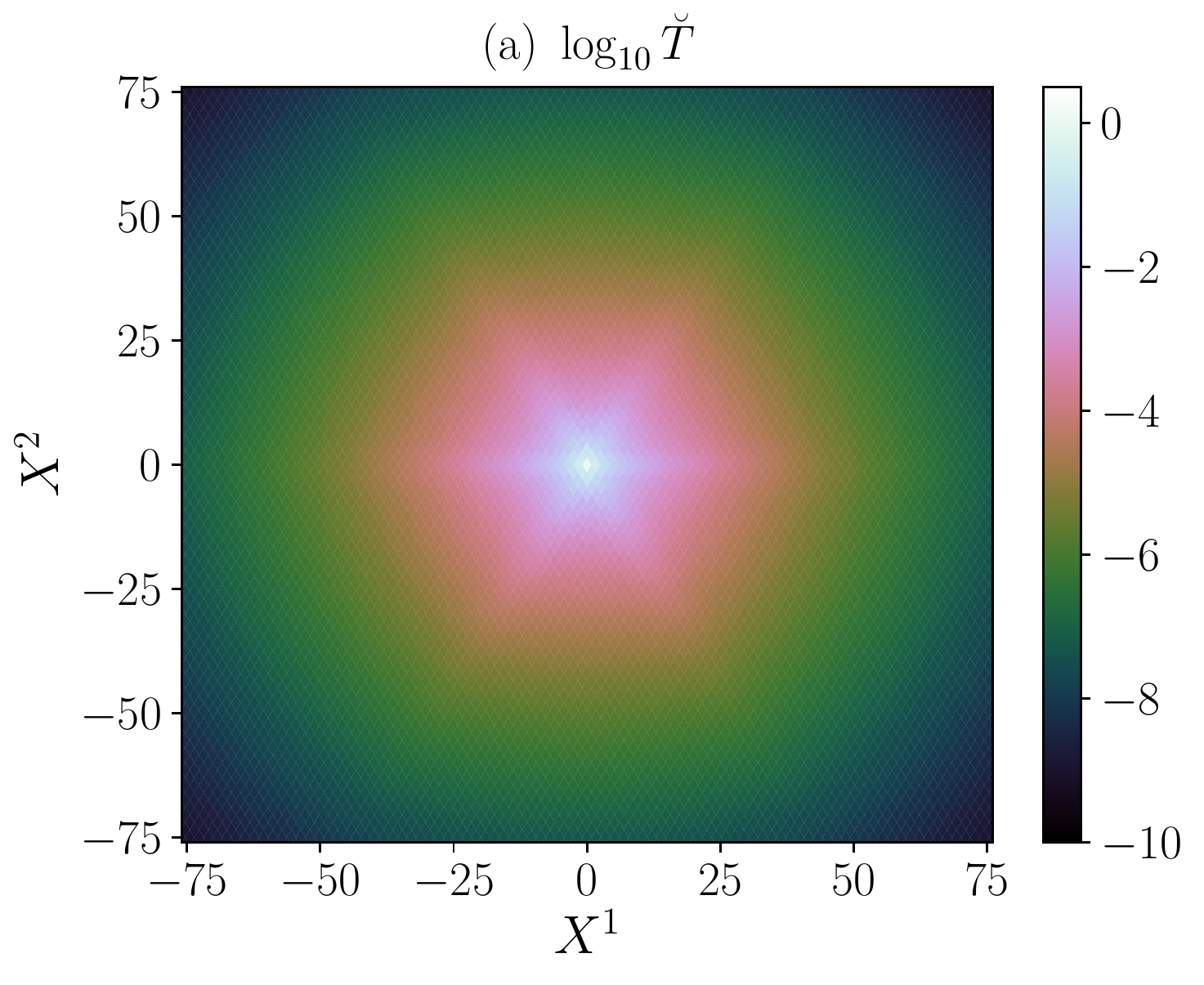}
\centering\includegraphics[width=0.6\columnwidth]{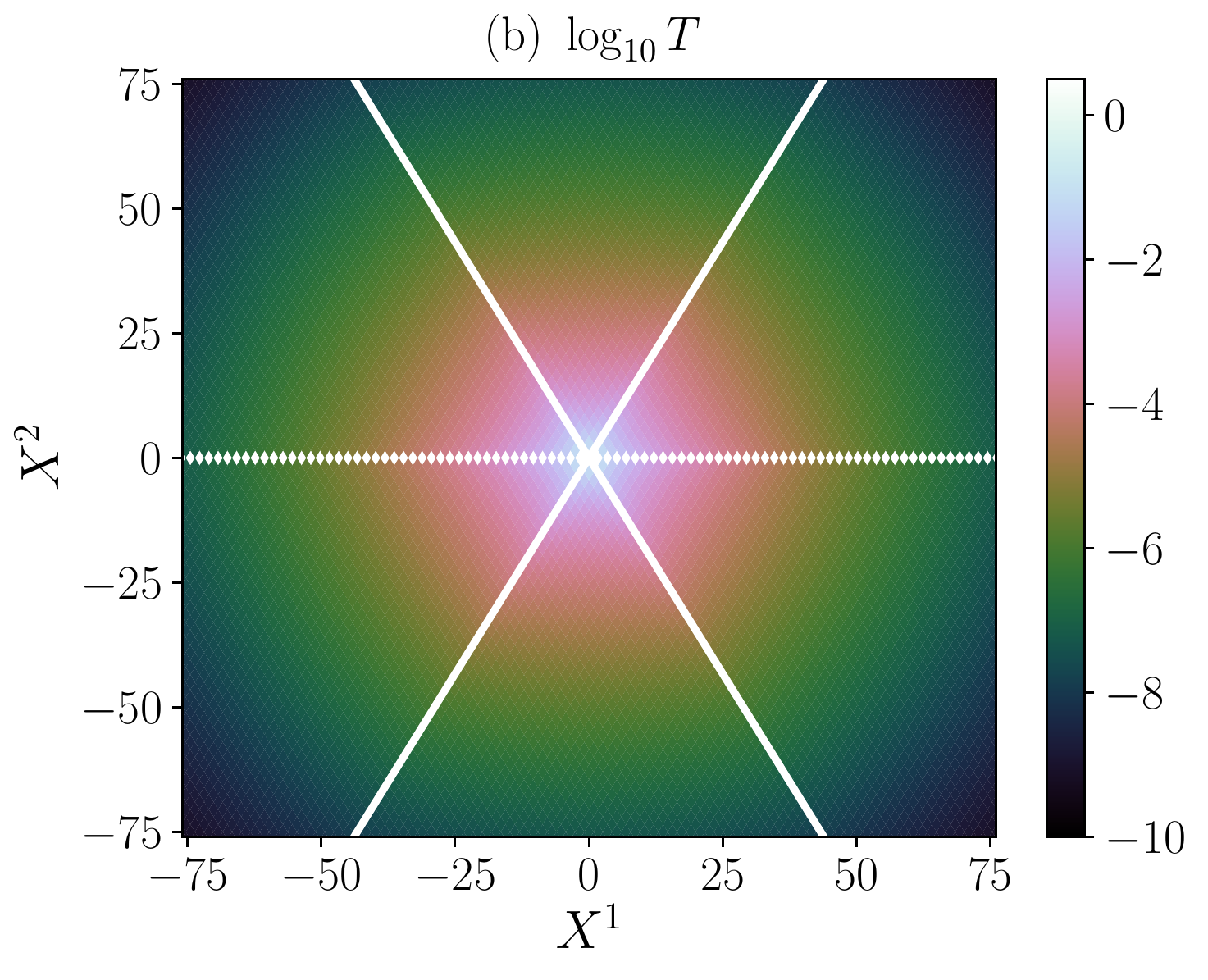}
\centering\includegraphics[width=0.6\columnwidth]{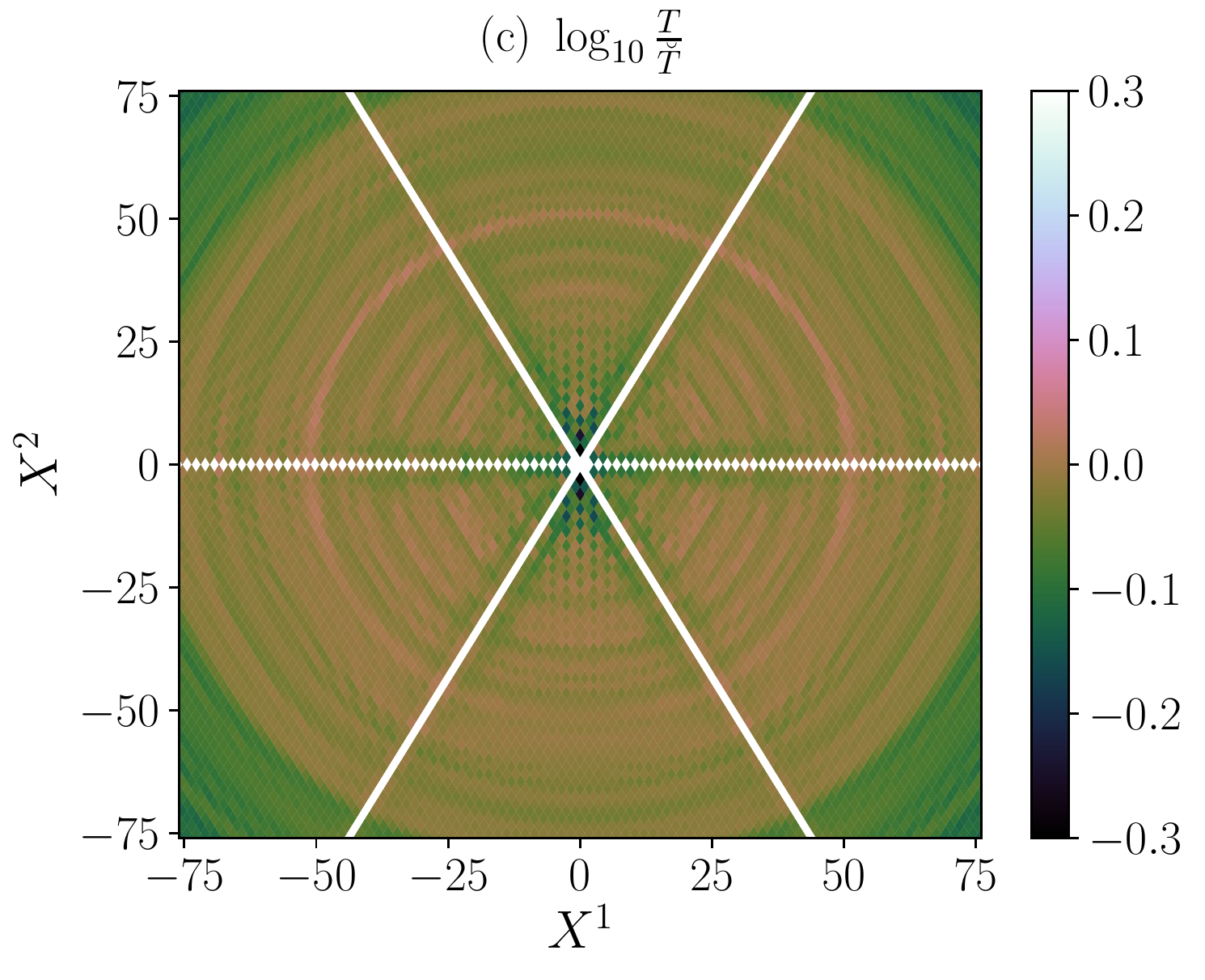}
\caption{2D plot for the kinetic temperature distribution pattern in the logarithmic scale: (a)
the discrete numerical solution $\breve T$,
(b) the analytical continuum solution $T$ (the singular
rays $\alpha
\in\left\{-\frac\pi3,0,\frac\pi3\right\} 
$
are shown by the white colour),
(c) the ratio $T/\breve T$.
}
\label{graphene-2d.pdf}
\end{figure}

\begin{figure}[p]	
\centering\includegraphics[width=0.73\columnwidth]{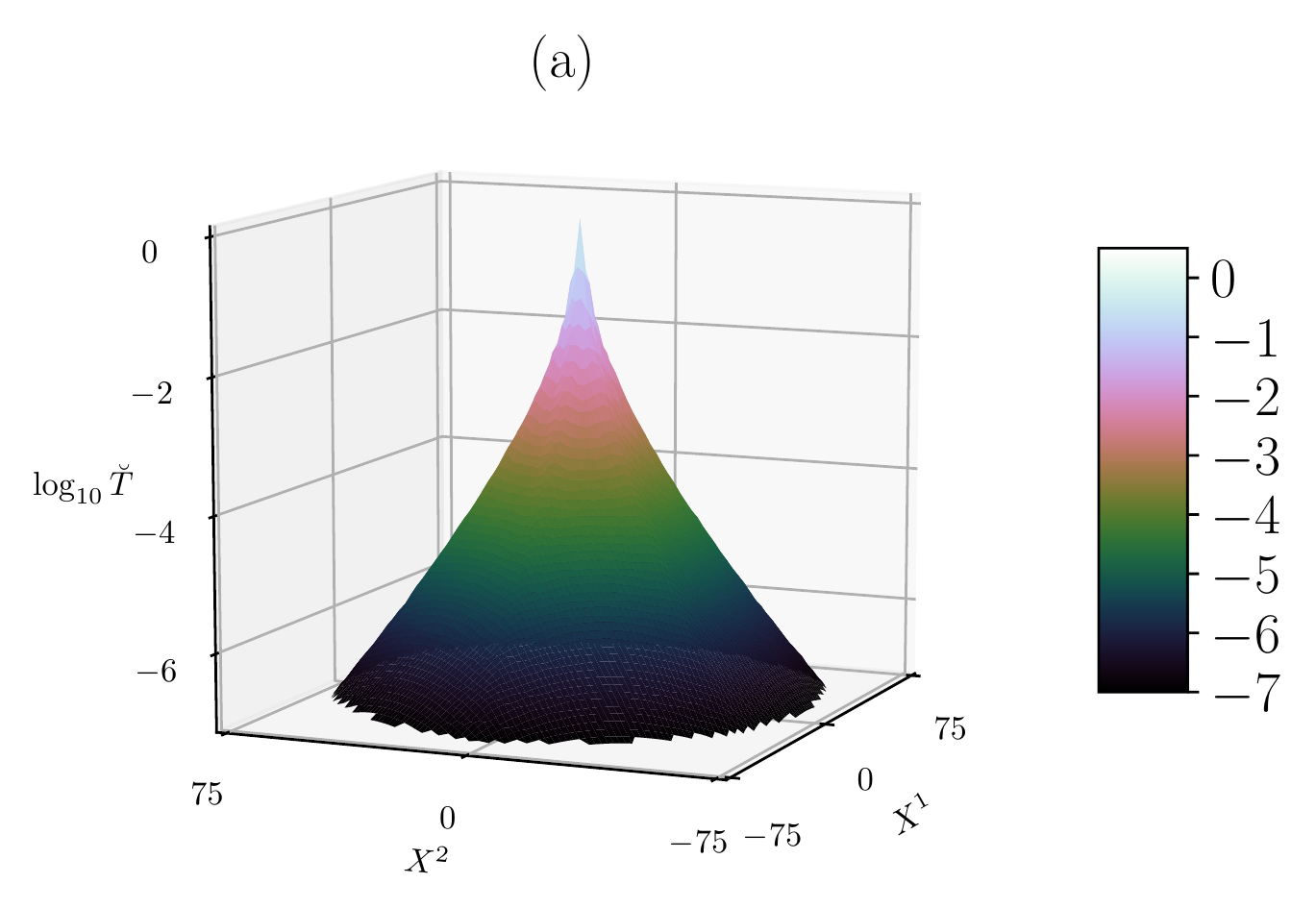}
\centering\includegraphics[width=0.73\columnwidth]{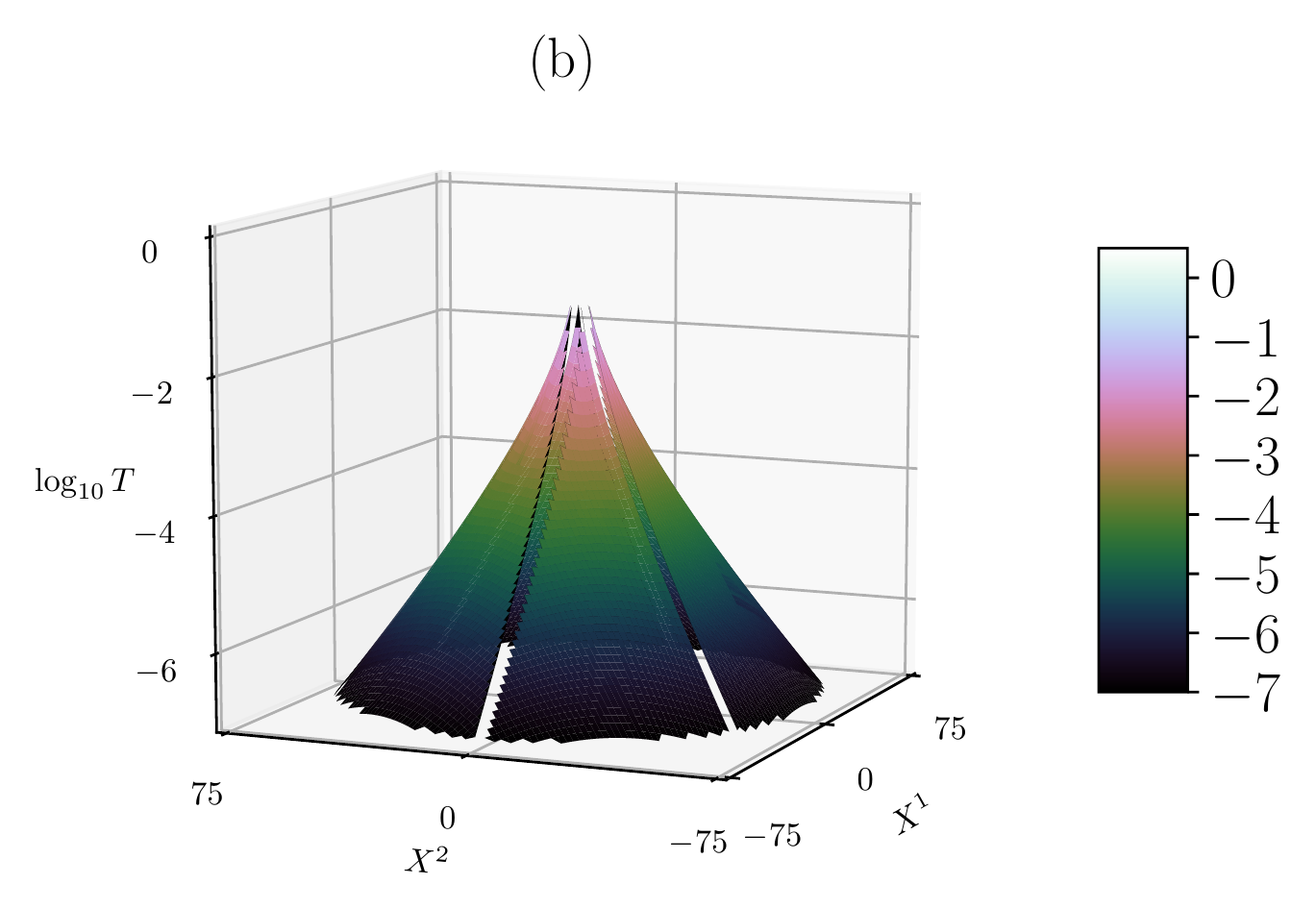}
\centering\includegraphics[width=0.73\columnwidth]{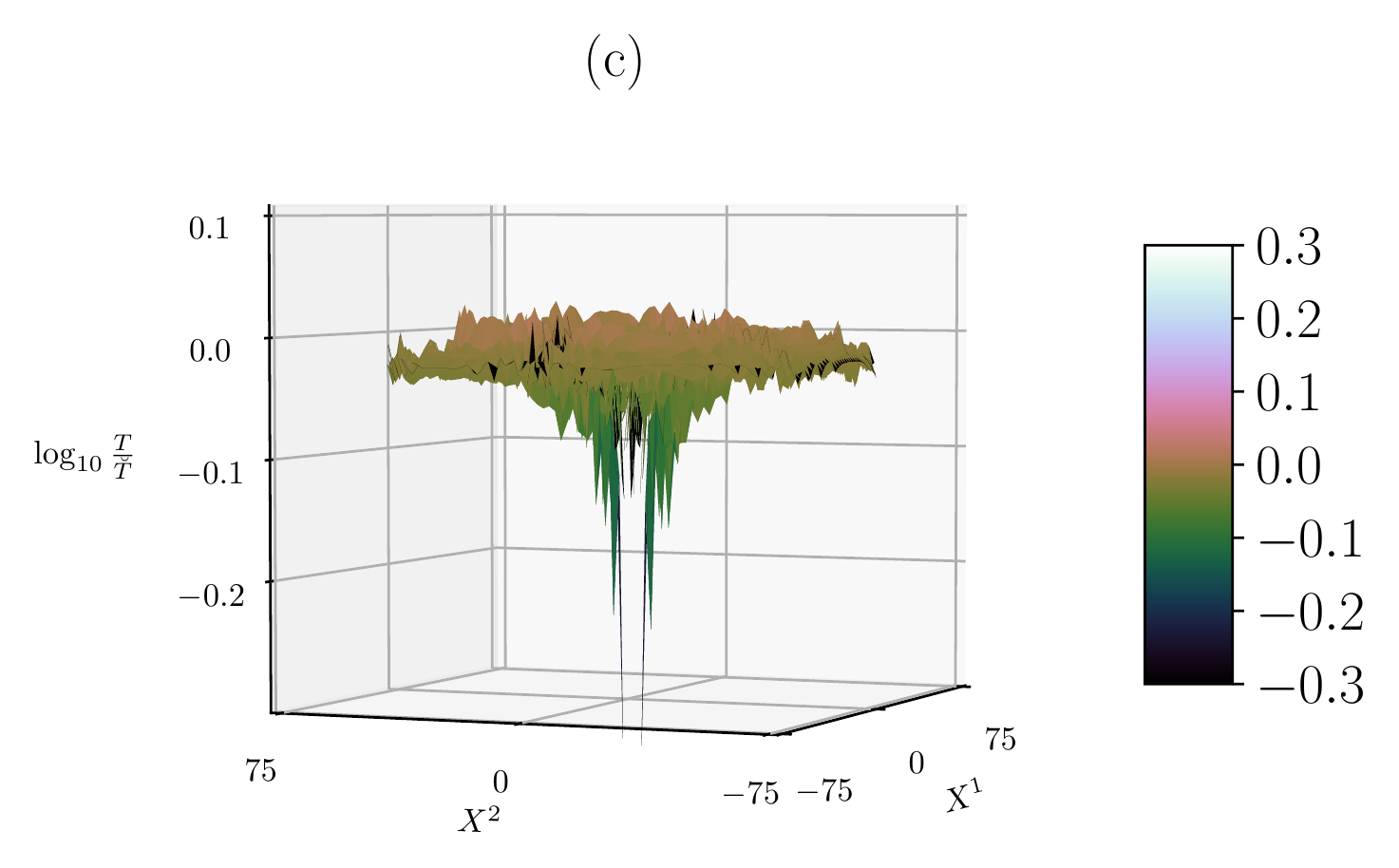}
\caption{3D plot for the kinetic temperature distribution pattern in the logarithmic scale: (a)
the discrete numerical solution $\breve T$,
(b) the analytical continuum solution $T$ (the singular
rays $\alpha
\in\left\{-\frac\pi3,0,\frac\pi3\right\} 
$
are shown as the transparent cuts),
(c) the ratio $T/\breve T$
}
\label{graphene-3d.pdf}
\end{figure}

\begin{figure}[p]	
\centering\includegraphics[width=0.75\columnwidth]{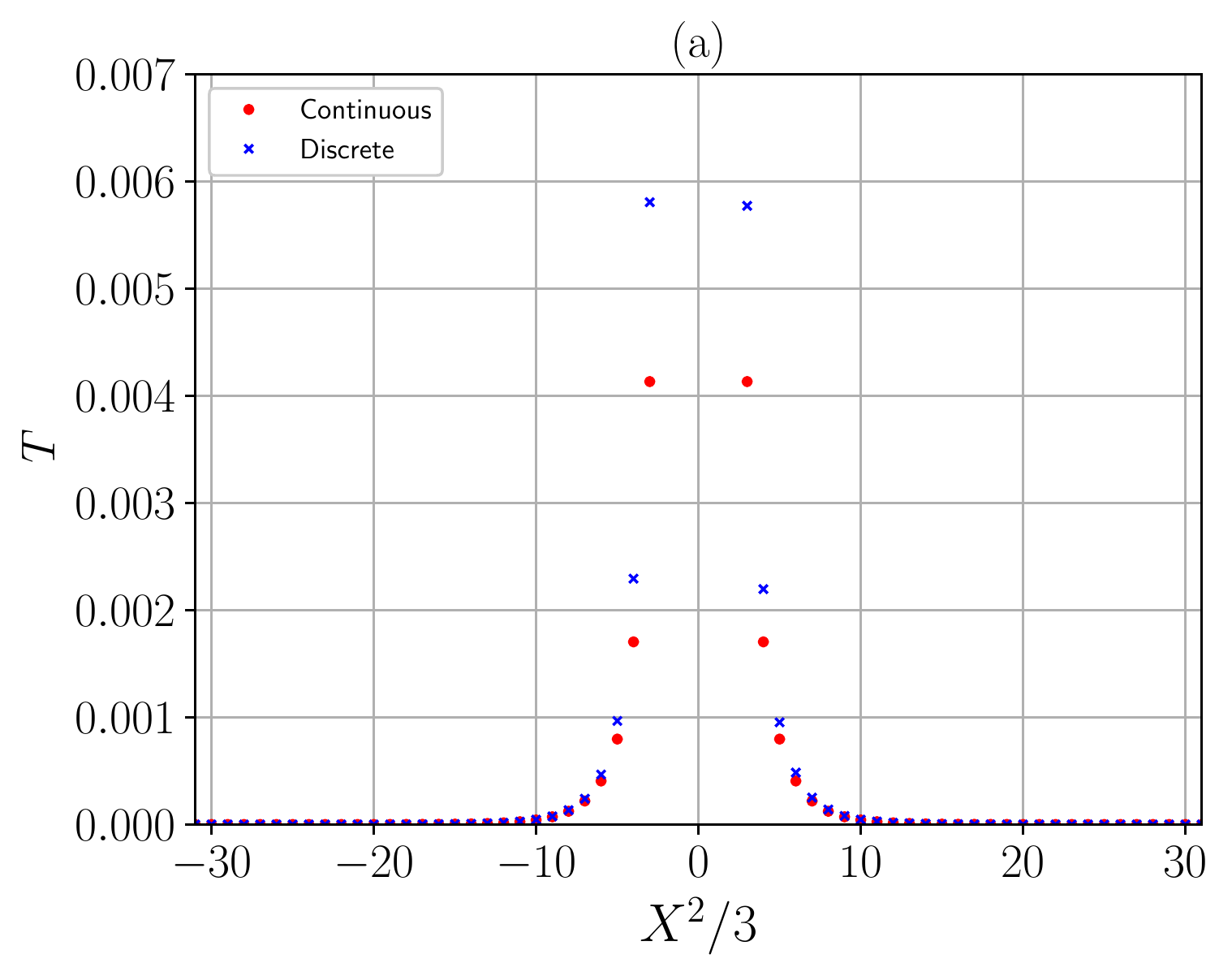}
\centering\includegraphics[width=0.75\columnwidth]{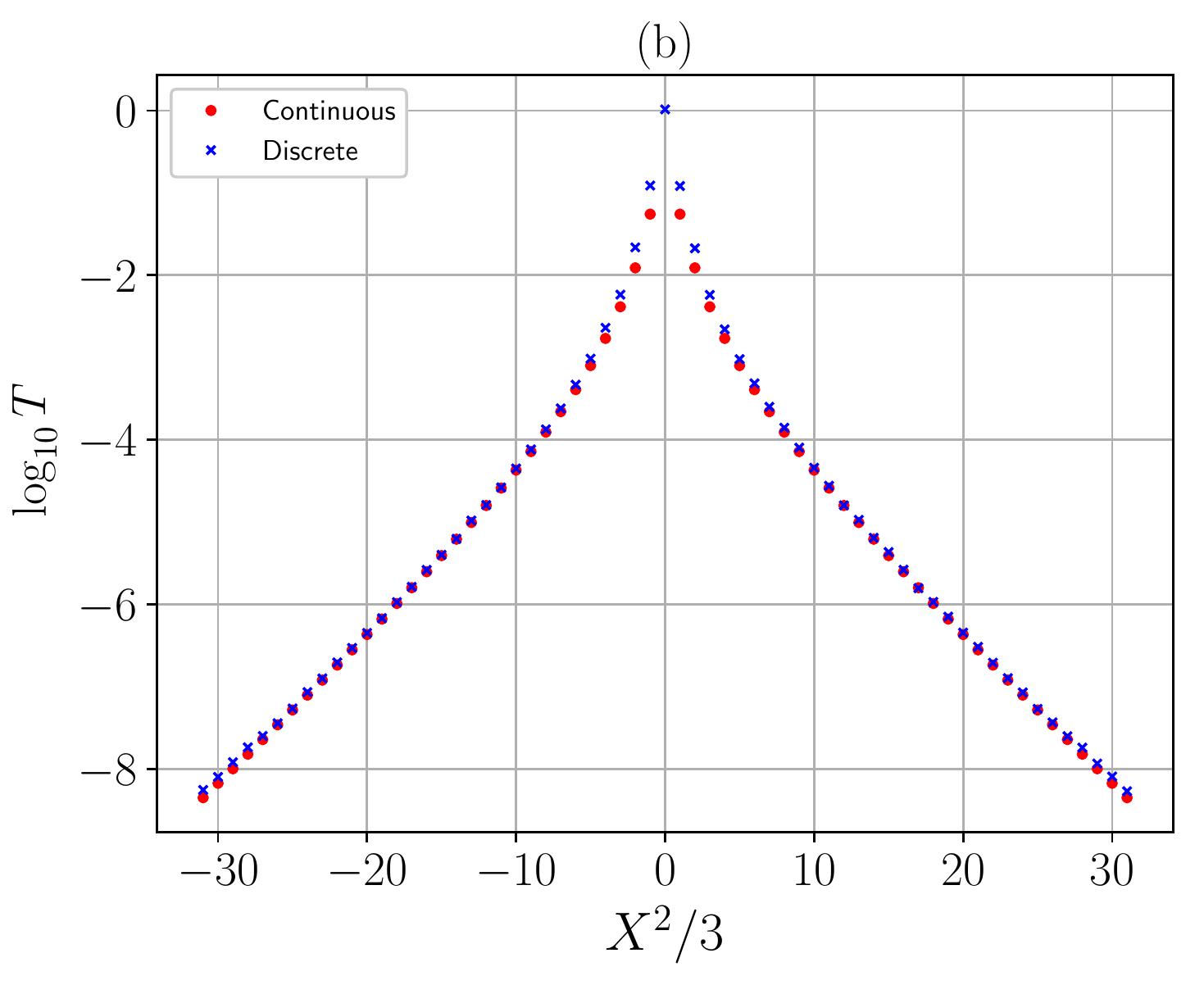}
\caption{Comparing the steady-state analytical continuum solution 
and the discrete
numerical solution 
for the row 
$\breve x^1=\breve x^2={X^2}/3\in\mathbb Z$
(along the armchair direction): (a) the linear scale, (b) the logarithmic scale}
\label{graphene-1d.pdf}
\end{figure}
\begin{figure}[tbp]	
\centering\includegraphics[width=0.75\columnwidth]{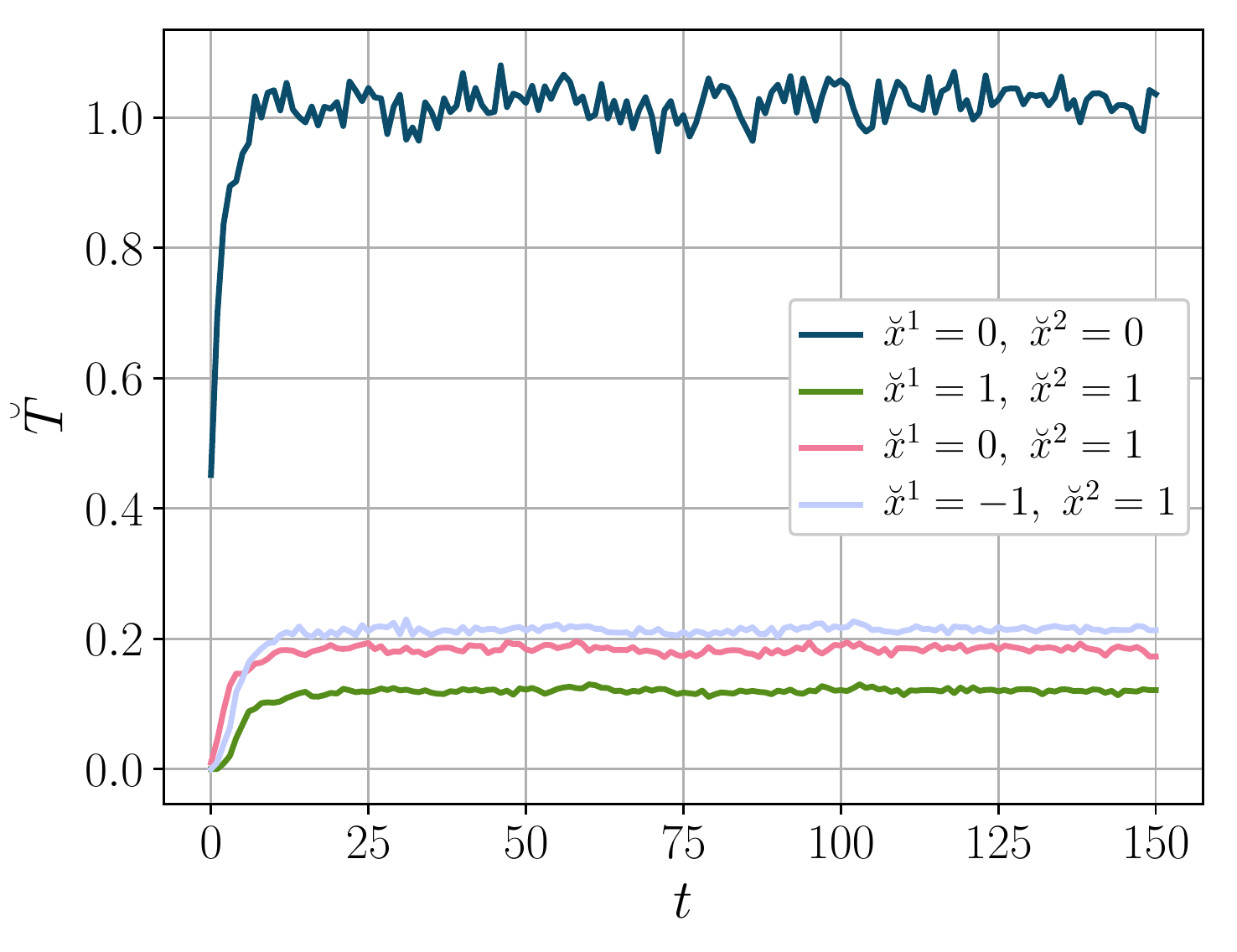}
\caption{Plots of the numerical discrete solution $\breve T$ at several
fixed positions versus the time (the plots are drawn with time step 1.0)}
\label{graphene-time.pdf}
\end{figure}

%
%
%

\section{Conclusion}
\label{gra-sec-conclusion}

In the paper we have generalized the results of recent papers 
\cite{gavrilov2018heat,gavrilov2019steady,Kuzkin-Krivtsov-accepted,Kuzkin2019}
and have suggested formula 
\eqref{gra-maineq-prealpha}, which allows us to give continuum description of
the steady-state
ballistic heat transport in the graphene sheet suspended under tension in viscous
media.
The obtained continuum solution corresponds to
the case where both graphene sub-lattices are equally excited 
by a heat source acting within a single primitive cell. According to \cite{Kuzkin2019},
we expect that the solution, in the 
case where only one particle of a single cell is
excited,
should have much more complicated structure.
The far-field of our solution, in principle, can be
compared with results of experiments with pure monocrystalline
graphene under laser heating. In such an experiment heat supply zone will
contain many primitive cells and, therefore, both sub-lattices will be
approximately equally excited. 

We expect that formula \eqref{gra-maineq-prealpha} is applicable for
various damped polyatomic lattices where all particles have
equal masses in
the case of universal (for all particles) external viscosity. As far as we
know for the time being, accurate
derivation of this formula is enough complicated problem, which can be a
subject for a separate work. The results of the present paper can be useful for
such a future study.
The obtained analytical continuum solution is in a good 
quantitative agreement
with the numerical discrete solution everywhere excepting a neighbourhood of
six singular rays \eqref{gra-singular-rays}
with the origin at the heat source location. The continuum
solution becomes singular at these rays, unlike the discrete one, which
appears to be localized in a certain sense along the rays. 
On the other hand, the demonstrated agreement between the discrete and the
continuum solution is a bit worse in
comparison with results previously obtained for a square scalar lattice
\cite{gavrilov2019steady}, especially in a zone near the heat source.
In our opinion, this fact is caused by the
continualization procedure used in \cite{Kuzkin2019}, which does not
take into account the particle arrangement inside a primitive cell (see
Remark~\ref{gra-remark-arra}). Indeed, near the heat source the continuum
solution is a function, which changes essentially inside a cell. 

In physical literature
the phenomenon of localization of the ballistic phonons along the singular
rays is known as the phonon focusing
\cite{Fu2020,Northrop1980,Northrop1979,Wolfe1998}.
This phenomenon is observed for both unsteady and steady-state problems and is commonly
interpreted in the framework of the stationary approach, where it is 
associated
\cite{Maris1971}
with the points on the dispersion surfaces, 
where condition \eqref{gra-Jacobian} is fulfilled.
In the case under consideration, such a
localization can be explained from an alternative point of view, 
which is based on the non-stationary conceptions  (see
Sect.~\ref{gra-sec-singular} and
\cite{slepyan1987energy,Abdukadirov_2019,AyzenbergStepanenko2008}). Namely, the
directional localization emerges due to the contribution from the resonant frequencies
\eqref{gra-resonant}
to which zero group velocity corresponds. All
of the energy contributed from such frequencies propagates along the singular
rays \eqref{gra-singular-rays} and forms the localized solutions. 

In order to
fix the mismatch on the
singular rays between the continuum and the discrete solutions, in our opinion,
one needs to modify the procedure of
continualization suggested in \cite{Kuzkin2019}. This can be done by constructing 
uniform asymptotics \cite{Fedoruk-Saddle,temme2014} in the framework of the method of
stationary phase, which can be used \cite{Gavrilov2020:2006.08197} to obtain the
continuum solution. Developing such a more accurate procedure of
continualization can be a subject of a future work. 
This will also explain in a reasonable way ``paradoxical'' behaviour of the
continuum solution on the singular rays, which was demonstrated in previous
papers \cite{gavrilov2019steady,Kuzkin-Krivtsov-accepted,Kuzkin2019}, where
steady and unsteady ballistic heat transfer is considered. 
The unsteady problem for an undamped square scalar lattice
considered in \cite{Kuzkin-Krivtsov-accepted} seems to be a simplest test problem for 
such a research.

\section*{Acknowledgements}
The authors are grateful to 
A.T.~Ivaschenko, V.A.~Kuzkin, O.V.~Gendelman, A.~Politi, E.V.~Shishkina, 
A.A.~Sokolov
for useful and stimulating
discussions. 

\section*{Funding}
This work
is supported by Russian Science Support Foundation (Grant No.~21-11-00378).

\appendix

\section{The dispersion surfaces and the group velocities for graphene lattice}
\label{gra-sec-app-dispersion}
Consider Eqs.~\eqref{gra-maineq},\eqref{gra-v-dot-u}
in the absence of dissipation ($\eta=0$) and
of the noise term in the right-hand side ($b_0=0$):
\begin{gather}
m\dt^2{\bm u}(\breve{\bb x})
+\bm C_0 \bm u(\breve{\bb x})
+\sum_{i=1}^{Q/2}
\left(
\bm C_1 \bm u(\breve{\bb x}+\bb b_i)
+\bm C_1^\top \bm u(\breve{\bb x}-\bb b_i)
\right)
=
\bb 0.
\label{gra-maineq-homo0}
\end{gather}
Let 
\begin{equation}
\bm u(\breve{\bb x})=\bm U\exp(-\I\bb p\cdot\breve{\bb x}-\I\omega t).
\label{gra-exp-r}
\end{equation}
Provided that $\omega$ satisfies the dispersion relations
\begin{equation}
\omega=\omega_\pm(p_1,p_2),
\end{equation}
the right-hand side of Eq.~\eqref{gra-exp-r} is the solution of
Eq.~\eqref{gra-maineq-homo0}. 
Here 
\begin{gather}
\bb p=p_\gamma\bb b^\gamma
\end{gather}
is the wave vector, $\bb b^\gamma$ is the dual basis defined by 
Eq.~\eqref{gra-bbe-dual}.


The dispersion relations for graphene lattice are found in
\cite{kuzkin2019thermal,Abdukadirov_2019,Ishibashi2014}: 
\begin{gather}
\omega_\pm^2=\omega_\ast^2(3\pm R(p_1,p_2)),\\
R(p_1,p_2)=\sqrt{3+2(\cos p_1 +\cos p_2 +\cos(p_1-p_2) )},
\label{gra-R-p1p2}
\\
\omega_\ast^2=\frac Cm.
\end{gather}
The plot of the dispersion surfaces is shown in Fig.~\ref{gra-omega-plot}.
\begin{figure}[htbp]	
\centering\includegraphics[scale=0.5]{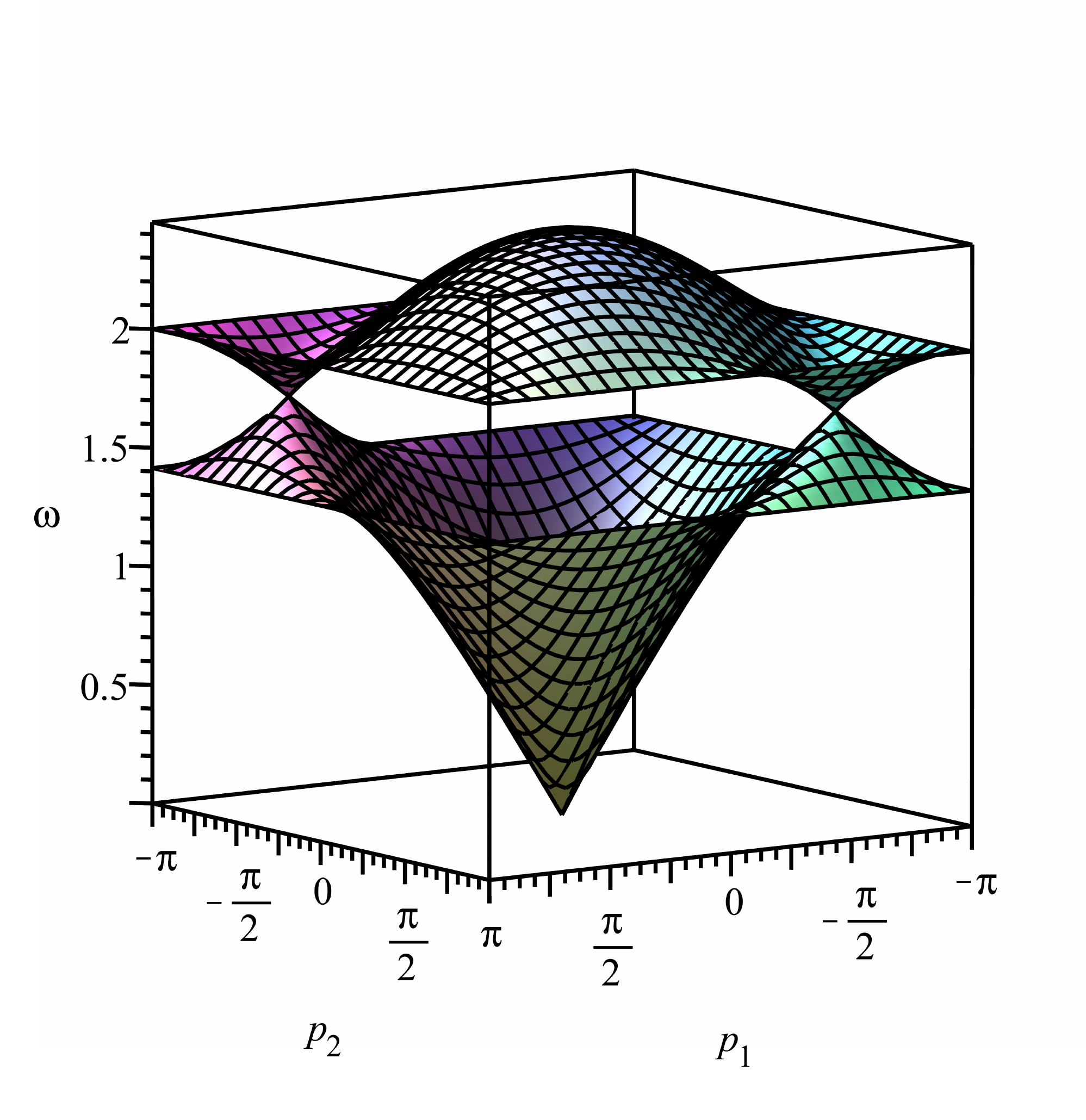}
\caption{The acoustic $\omega_-(p_1,p_2)$ and the optic $\omega_+(p_1,p_2)$
dispersion surfaces ($\omega_\ast=1$)}
\label{gra-omega-plot}
\end{figure}

The vectors of the group velocities are
\begin{gather}
{\groopv}_\pm=
\pd{\omega_\pm}{\bb p}
=
\pd{\omega_\pm{}}{p_\gamma}\bb b_\gamma,
\label{gra-C-def}
\\
g^1_\pm=\pd{\omega_\pm}{p_1}
=\mp\frac{\omega_\ast^2(\sin p_1+\sin(p_1-p_2))}{2\omega_\pm R},
\label{gra-C1}
\\
g^2_\pm=\pd{\omega_\pm}{p_2}
=\mp\frac{\omega_\ast^2(\sin p_2-\sin(p_1-p_2))}{2\omega_\pm R},
\label{gra-C2}
\\
\bC_\pm=\mp\frac{\sqrt3c\big(
(2\sin(p_1-p_2)+\sin p_1-\sin p_2)\bb e_1
+\sqrt3(\sin p_1+\sin p_2)\bb e_2
\big)}
{4R\sqrt{3\pm R}}.
\label{gra-group-initial}
\end{gather}
Here 
\begin{equation}
c=\omega_\ast a
\label{gra-c-def}
\end{equation}
is a characteristic speed.
\bibliographystyle{plain}
\bibliography{math,thermo,serge-gost,discrete,graphene,web,moving_load}

\selectlanguage{english}

\end{document}